\documentclass[10pt,prd,singlecolumn,
nofootinbib, noshowkeys, noshowpacs, superscriptaddress, floatfix, aps]{revtex4-2}
\usepackage{amsmath,amsfonts,amsthm,amssymb}
\usepackage[dvips]{graphics,graphicx}
\usepackage[usenames,dvipsnames]{xcolor}
\definecolor{darkblue}{RGB}{0,0,196}
\definecolor{darkgreen}{RGB}{0,120,0}
\usepackage[colorlinks=true,linktocpage=true,linkcolor=darkblue,citecolor=red,urlcolor=darkblue]{hyperref}
\usepackage{hyperref}
\usepackage[export]{adjustbox}
\usepackage{tikz}
\usepackage{appendix}
\usepackage{pifont}
\usetikzlibrary {arrows.meta}
\usepackage[mathscr]{euscript}
\usepackage[export]{adjustbox}
\newcommand{\no}{\nonumber}
\newcommand\hstar[1]{\ThisStyle{\ensurestackMath{%
\setbox0=\hbox{$\SavedStyle#1$}%
\stackengine{0pt}{\copy0}{\kern.2\ht0\smash{\SavedStyle\star}}{O}{c}{F}{T}{S}}}}
\definecolor {darkgreen}{rgb}{0.2,0.7,0.2}
\begin{document}
\title{Kinetic Theory of Quasiparticles, Retarded Correlators and Hydrodynamics}
\author{Ali Hataei}
\affiliation{Department of Physics, Shahid Beheshti University, Tehran, Iran}
\email{a.hataei@alumni.ac.ir}
\author{Roya Heydari}
\email{rheydari@ipm.ir}
\author{Farid Taghinavaz}
\email{ftaghinavaz@ipm.ir}
\affiliation{School of Particles and Accelerator, Institute for Research in Fundamental Sciences (IPM), P. O. Box 19395-5531, Tehran, Iran.}
\begin{abstract}
Within the relaxation time approximation under a constant mass profile, we investigate the collective dynamics of a system of massive relativistic particles described by the Maxwell-Boltzmann equilibrium distribution. We analytically derive the two-point retarded correlation functions for both charge and energy-momentum tensor components at arbitrary momentum and frequency. We expand our results in the limits of very small and very large mass-to-temperature ratios ($m/T$). Similar to the massless case, we identify a critical threshold in ($k \tau$) below which the correlators permit physical solutions. This behavior arises from a logarithmic branch cut in the spectral function. At higher momenta, solutions emerge significantly below this cut, corresponding to non-hydrodynamic modes. Our analysis demonstrates that hydrodynamic poles dominate in the strong coupling regime, while the weak coupling regime features a logarithmic branch cut extending along $\omega = k$ and $\omega = -k$. Notably, in the sound channel, finite mass modifies the standard propagating sound mode, converting it into a purely imaginary mode. In contrast, the shear channel exhibits modes that asymptotically converge to their massless counterparts. Additionally, we compute the transport coefficients for shear and bulk viscosity, along with higher-order gradient corrections up to third order, expressed as perturbative expansions in both the small and large ($m/T$) regimes.
\end{abstract}

\maketitle
\section{Introduction}
Extensive evidence has confirmed that a new phase of quark matter, known as the Quark-Gluon Plasma (QGP), emerges after heavy-ion collisions. This phase exists in a state of nearly local thermal equilibrium and is characterized by almost free and deconfined quarks and gluons, which exhibit perfect fluid-like behavior at macroscopic scales \cite{PHENIX:2004vcz, Gyulassy:2004zy, Teaney:2000cw, Rischke:2003mt, Shuryak:2004cy}. Similar collective behavior has also been observed in collisions involving smaller hadrons \cite{CMS:2010ifv, CMS:2016fnw}. To describe the bulk dynamics in such systems, Relativistic Hydrodynamics (RH) is an acceptable framework \cite{Kovtun:2012rj, Florkowski:2017olj, Romatschke:2017ejr, Rezzolla:2013dea}. This approach traditionally relies on two key assumptions: (1) having a local thermal equilibrium state, and (2)  distinct separation of scales between microscopic and macroscopic dynamics. The consequence of the latter condition is usually cast in terms of a small Knudsen number series, ensuring that a regular gradient expansion can describe all hydrodynamic quantities. However, the observation of collective flow in small hadron collisions has challenged this conventional paradigm \cite{Noronha-Hostler:2015wft, Heinz:2019dbd, Weller:2017tsr}.

Although RH exhibits remarkable power in describing bulk properties, it is confronted with several unresolved questions \cite{Busza:2018rrf}. One significant limitation is its inability to elucidate the role of microscopic processes, such as particle collisions or the strength of coupling between particles. To address this issue, two different approaches can be employed: (1) gauge/gravity duality and (2) Kinetic Theory. 

The first approach is the gauge/gravity duality or the AdS/CFT correspondence as a specific example of this duality that relates the dynamics of strongly coupled superconformal field theories on the boundary of asymptotically Anti-de Sitter (AdS) spacetime to the bulk dynamics of a superstring model in the large 't Hooft coupling limit \cite{Maldacena:1997re, Aharony:1999ti}. Within this paradigm, numerous near-equilibrium and out-of-equilibrium properties of a strongly coupled QCD plasma can be inferred. These include shear viscosity \cite{Kovtun:2004de, Policastro:2001yc}, second-order transport coefficients \cite{Baier:2007ix}, the energy loss of heavy quarks traversing in a $\mathcal{N}=4$ Yang-Mills plasma \cite{Herzog:2006gh, Casalderrey-Solana:2006fio}, quasi-normal modes corresponding to long-lived modes of the boundary theory \cite{Kovtun:2005ev}, and the calculation of jet quenching parameters \cite{Liu:2006ug}, and so on. 

The second approach is the kinetic theory, which provides a more suitable framework for addressing limitations of the RH \cite{DeGroot:1980dk}. Kinetic theory is based on two primary assumptions: (1) a rarefied system with weak interactions, and (2) the hypothesis of irreversible collisions, commonly referred to as the "molecular chaos" hypothesis. Within this framework, the evolution of the one-particle distribution function in phase space is governed by the Boltzmann equation. By expressing macroscopic quantities, such as the energy-momentum tensor or current vector components, as appropriate moments of this distribution function, the dynamical behavior of these quantities can be determined.

At first glance, the kinetic theory with its foundational assumptions may appear incapable of reproducing collective behavior. However, this is not the case. Many years ago, it was demonstrated that approaching a local thermal equilibrium in a boost invariant system is possible by using the relativistic kinetic theory within the relaxation-time approximation \cite{Baym:1984np}. In recent years, significant efforts have been made to elucidate that kinetic theory, when coupled to a suitable interaction kernel, serves as a robust framework for bridging the microscopic and macroscopic regimes. For instance, it has been shown that kinetic theory employing the relaxation-time approximation can reproduce hydrodynamic modes across various scenarios through an appropriate choice of momentum and relaxation time parameters \cite{Romatschke:2015gic, Kurkela:2017xis, Bajec:2024jez, Abbasi:2024pwz, Brants:2024wrx, Ochsenfeld:2023wxz, Kamata:2020mka}. Moreover, recent work has proposed that within the BBGKY hierarchy of density matrix evolution, a reasonable truncation of the lowest-order equation can reveal collective modes in the spectra of conserved current correlations \cite{Grozdanov:2024fxr}. Additionally, numerical computations have emphasized the value of kinetic theory in exploring both hydrodynamic and non-hydrodynamic modes, as well as flow observables. A widely adopted method for solving the Boltzmann equation numerically is the relativistic lattice Boltzmann algorithm \cite{Romatschke:2011hm}. Using this approach, it has been demonstrated that for a boost-invariant conformal system with particle-like excitations, the relative contributions of fluid dynamical and particle-like excitations can be inferred from the zeroth and higher-order moments of viscous fluid dynamics \cite{Kurkela:2019kip}. Similarly, numerical simulations of kinetic theory in this framework have successfully interpreted the large azimuthal anisotropy observed in proton-proton and proton-nucleus collisions \cite{Kurkela:2018qeb}. The development of numerical methods for systems of arbitrary size and transverse geometry has proven particularly useful for studying azimuthal flow coefficients, including non-linear mode-mode coupling, and for incorporating realistic event-by-event fluctuations in initial conditions \cite{Kurkela:2020wwb}. The kinetic theory also provides valuable insights into the harmonic flow response to initial geometric deformations in small systems through parametric scaling relations \cite{Kurkela:2021ctp, Taghavi:2019mqz}. Importantly, the kinetic theory offers critical perspectives on the limitations of hydrodynamic models. For instance, collective flow in high-energy collisions, as derived from numerical solutions of the Boltzmann equation, is an opacity-dependent quantity. This dependence can either validate or challenge the use of RH in interpreting flow observables in both small and large systems \cite{Ambrus:2022koq, Ambrus:2022qya}. Another application of numerical computations in the kinetic theory is the investigation of conditions under which transverse flow develops in collisions of both small and large systems \cite{Ambrus:2021fej}. These numerical simulations can be extended to other collision systems, such as oxygen+oxygen or gold+gold, to quantify the extent to which a macroscopic hydrodynamic description accurately captures the development of collective flow \cite{Ambrus:2024hks, Ambrus:2024eqa}. They also help to determine the sensitivity of collective flow in small systems to the non-equilibrium evolution of the QGP beyond hydrodynamic approximations \cite{Duguet:2025hwi}.

Motivated by these considerations, it is crucial to explore collective behavior from a microscopic framework such as the kinetic theory. A critical gap in this endeavor is the investigation of hydrodynamic behaviors in systems of massive particles. This investigation is vital for two key reasons. First, the introduction of a mass scale breaks conformal symmetry, bringing the analysis closer to realistic physical scenarios. Second, in the context of phase transition arguments within effective field theories, mass serves as an order parameter for chiral symmetry restoration/breaking, and its running can help to identify the location of critical points. Consequently, studying the collective behavior of a massive system using the kinetic theory would provide valuable insights into the validity of RH near critical points. Such an approach could reveal the extent to which RH remains applicable near the critical points. The kinetic theory of massive particles, often referred to as quasiparticles, has been explored in several contexts. For example, it has been employed to derive the equations of second-order relativistic viscous hydrodynamics for a system of temperature-dependent quasiparticles \cite{Tinti:2016bav}. Additionally, the method of moments has been applied to investigate hydrodynamization in a one-dimensional boost-invariant plasma with finite particle mass \cite{Jaiswal:2022udf}. Furthermore, transport coefficients in second-order transient hydrodynamics have been analyzed within this framework \cite{Rocha:2024rce}.

In this paper, we investigate the Boltzmann equation within the Relaxation Time Approximation (RTA) for a system of particles with a constant mass profile. The chemical potential of the particles is neglected in the background limit. Employing a variational approach, we derive retarded two-point correlation functions for the charge and energy-momentum tensor components at arbitrary $(\omega, k)$. To explore collective behaviors, we examine the small-momentum limits of these functions. For the sake of analytical tractability, we expand expressions for both small and large values of the mass-to-temperature ratio, \( x = m/T \). We obtain collective modes as well as the diffusion constant for extreme $x$ limits in the current-current correlation function.
In the energy-momentum sector, we derive the fluctuations of the energy-momentum components and the complete spectrum of retarded correlation functions. Up to certain contact terms, the Ward identities are satisfied. Furthermore, we compute the shear and bulk transport coefficients as well as higher-order transports in derivatives for both small and large $x$ values. 
The hydrodynamic poles in the shear and sound channels are identified, revealing that in the sound channel, all modes become purely imaginary. 
Additionally, we observe the emergence of a branch cut between the $\omega = k$ and $\omega = - k$ in the solutions in the weak coupling regime ($\tau_{eq} T \to \infty$), while in the strong coupling regime ($\tau_{eq} T \to \infty$), hydrodynamic poles arises above a critical $(k \tau_{eq})$. This is the consequence of the logarithmic branch cut for small momenta. 

The organization of this paper is as follows. In Section II, we review the Boltzmann equation, outline the methodology for obtaining retarded two-point functions, and discuss the thermodynamics of a massive Maxwell-Boltzmann distribution. In Section III, we analyze the correlation functions in the charge sector, examining their behavior for both small and large values of \( x \). Retarded correlators and hydrodynamic limit of these function are also studied. In Section IV, we investigate the energy-momentum fluctuations near the global equilibrium state and explore their hydrodynamic properties. Following this, we provide a brief discussion on the analytical behavior of the correlation functions in both the charge and energy-momentum sectors. Finally, we conclude our discussion with a summary of our findings and an outlook for future research. Two appendices are included to provide detailed derivations and computational steps.
\section{Massive Boltzmann equation and retarded correlators}\label{sec: thermo}
The Boltzmann equation describes the evolution of a one-particle distribution function \( f(t, \Vec{x}, \Vec{p}) \) in phase space \((\Vec{x}, \Vec{p})\). Solving the Boltzmann equation with a general collision kernel is highly complex. However, it becomes significantly more tractable under certain approximations. One such approximation is the well-known Relaxation-Time Approximation (RTA), referred to as the Bhatnagar-Gross-Krook model in the classical formulation \cite{Bhatnagar:1954zz}. In a relativistic framework, this approximation is described by the Anderson-Witting model \cite{Anderson:1974nyl}. We adopt the relativistic version of the RTA, which can be expressed as follows
\begin{align}\label{eq:eq-BE}
    p^{\nu}\partial_{\nu}f(t,\Vec{x}, \Vec{p}) + F^{\alpha}\partial_{\alpha}^{(p)}f(t,\Vec{x}, \Vec{p}) = - \frac{p^{\alpha}u_{\alpha}}{\tau_{eq}} \left(f(t, \Vec{x}, \Vec{p}) - f_{eq}(t, \Vec{x}, \Vec{p})\right).
\end{align}
The first term on the left-hand side describes the evolution in space-time coordinates, while the second term, coupled to a force term \( F^\alpha \), represents the evolution in momentum space, where \( \partial^{(p)}_\alpha = \frac{\partial}{\partial p_\alpha} \).  
\footnote{Throughout this paper, we use the notation \(\partial_\mu \equiv \frac{\partial}{\partial x^\mu}\). Any deviations from this convention will be explicitly stated.}  
Here, \( u^\alpha \) denotes the average velocity of the particles, and \( \tau_{eq} \) is the equilibrium time scale over which the distribution function evolves toward its equilibrium distribution \( f_{eq}(t, \Vec{x}, \Vec{p}) \). In the current paper, we neglect quantum statistics and choose the equilibrium Maxwell-Boltzmann distribution function  \cite{Romatschke:2015gic}
\begin{align}  
f_{eq}(t, \Vec{x}, \Vec{p}) = \exp{\left(\frac{g_{\alpha \beta} p^\alpha u^\beta(t, \Vec{x}) + \mu(t, \Vec{x})}{T} \right)}.
\end{align}  
Having the solutions to Eq. \eqref{eq:eq-BE} enables us to derive the energy-momentum tensor and charge current components in the following forms
\begin{align}\label{eq: def-T- J}
    &T^{\alpha \beta}(t, \Vec{x}) = \int \, \frac{d^3p}{(2\pi)^3} \, \frac{p^\alpha p^\beta}{p^0} f(t, \Vec{x}, \Vec{p}), \no\\
    & J^{\alpha}(t, \Vec{x}) = q \int \, \frac{d^3p}{(2\pi)^3} \, \frac{p^\alpha}{p^0} f(t, \Vec{x}, \Vec{p}),
\end{align}
where \( q \) is the (electric) charge of a single particle. The variables on the left-hand side operate on the macroscopic scale and can be matched to RH constitutive relations. These macroscopic quantities are described by integrating over microscopic-scale functions such as \( f(t, \Vec{x}, \Vec{p}) \).

The force term may have different contributions. External electromagnetic fields, the curvature of space-time, and a varying mass profile can serve as sources for it \cite{Florkowski:1995ei, Romatschke:2011qp}
\begin{align}\label{eq: term-F}
    F^\alpha =  F^\alpha_{(em)} + F^\alpha_{(gr)} + M \partial^\alpha M.
\end{align}
For our purposes, where we assume constant mass values, the first term on the right-hand side does not contribute. We postpone the study of systems with varying mass to future work. The remaining two contributions take the following forms
\begin{align}\label{eq: term-F-em-g}
    & F^{\alpha}_{(em)} = F^{\alpha \beta} p_\beta = (\partial^\alpha A^\beta - \partial^\beta A^\alpha)p_\beta, \no\\
    & F^\alpha_{(gr)} = - \Gamma^\alpha_{\beta \gamma} p^\beta p^\gamma =-\frac{1}{2} g^{\alpha \rho} \left( \partial_\beta g_{\rho \gamma} + \partial_\gamma g_{\rho \beta} - \partial_\rho g_{\beta \gamma}\right) p^\beta p^\gamma.
\end{align}
Here, \( F^{\alpha \beta} \) and \( \Gamma^\alpha_{\beta \gamma} \) are the electromagnetic field-strength tensor and the Christoffel symbol, respectively.
\subsection{Retarded Correlator}
The retarded correlation functions of the charged current and the energy-momentum tensor are typically defined as follows
\begin{align}
    &G_{\mu \nu}(t-t', \Vec{x} - \Vec{x}') = - i \theta(t - t') \langle \left[ J_\mu(t, \Vec{x}), J_\nu(t', \Vec{x}')\right]\rangle_{eq}, \no\\
    &G_{\mu \nu, \alpha \beta}(t-t', \Vec{x} - \Vec{x}') = - i \theta(t - t') \langle \left[ T_{\mu \nu}(t, \Vec{x}), T_{\alpha \beta}(t', \Vec{x}')\right]\rangle_{eq}.
\end{align}
Here, the averages are taken with respect to the thermal equilibrium state. There are two methods to calculate these correlation functions: (1) the canonical approach and (2) the variational approach.

In the canonical approach, linear response theory is employed. Sources corresponding to conserved quantities, such as particle number and energy-momentum, are introduced adiabatically, and the response of hydrodynamic variables to these weak perturbations is computed at first order \cite{Kovtun:2012rj}. In contrast, the variational approach involves coupling sources directly to their conjugate conserved operators. For example, the gauge field \(\delta A_\mu(t, \Vec{x})\) couples to the (electric) current, while the metric field \(\delta g_{\mu \nu}(t, \Vec{x})\) couples to the energy-momentum tensor. In this framework, the retarded correlators are defined as
\begin{align}\label{eq: def-var}
    &G_{\mu \nu}(t-t', \Vec{x} - \Vec{x}') = - \frac{\delta \mathcal{J}_\mu(t, \Vec{x})}{\delta A^\nu(t', \Vec{x}')}\bigg|_{\delta A = \delta g = 0}, \no\\
    &G_{\mu \nu, \alpha \beta}(t-t', \Vec{x} - \Vec{x}') = - 2 \frac{\delta \mathcal{T}_{\mu \nu}(t, \Vec{x})}{\delta g^{\alpha \beta}(t', \Vec{x}')}\bigg|_{\delta A = \delta g = 0},
\end{align}
where 
\begin{align}
    & \mathcal{J}_\mu = \sqrt{-g} \, J_\mu, \no\\
    & \mathcal{T}_{\mu \nu} = \sqrt{-g} \, T_{\mu \nu}.
\end{align}
The presence of the \(\sqrt{-g}\) term introduces a contact term through the relation \(\delta \sqrt{-g} = \frac{1}{2} \sqrt{-g} g^{\mu \nu} \delta g_{\mu \nu}\).

We adopt the variational approach to compute the correlation functions in this work. For simplicity, we neglect the contact terms, and the derivatives in Eq. \eqref{eq: def-var} are taken for the original current and energy-momentum tensor components.
\subsection{Thermodynamics of massive Maxwell-Boltzmann particles}
\begin{figure}
    \centering
    \includegraphics[width=0.485\textwidth]{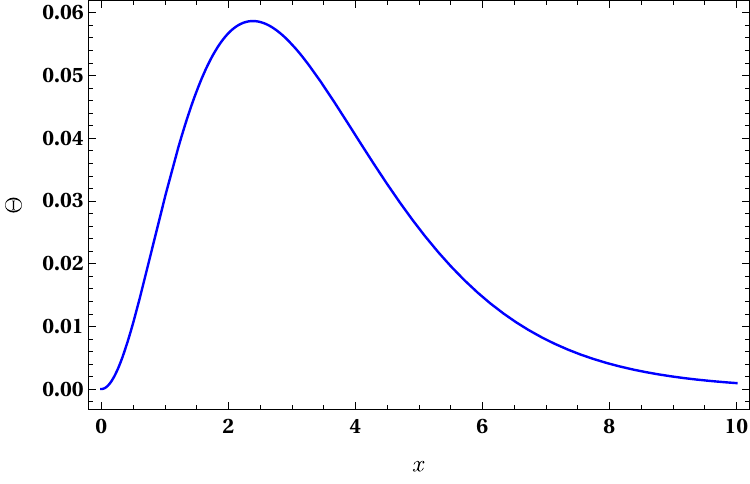}
\hspace{0.3cm}
    \includegraphics[width=0.485\textwidth]{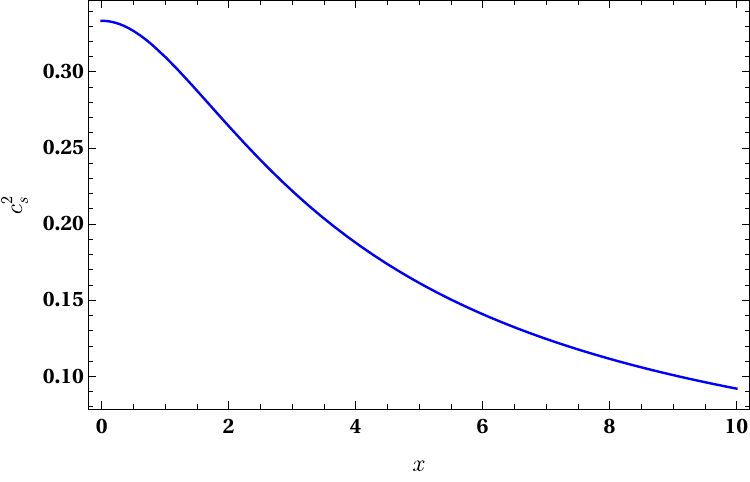}
\caption{Left: the trace anomaly parameter $\Theta$ as given in Eq. \eqref{eq:trace-cs2}, and right: the speed of sound $c_s^2$ as functions of $x = m/T$.}
    \label{fig: plot-cs2-trace}
\end{figure}

In this short part, we review the thermodynamic quantities of massive Maxwell-Boltzmann particles. The equilibrium distribution function is given by 
\begin{align}\label{eq: equ-MB}
    f_{eq} = \exp\left(-\frac{p_0 - \mu_0}{T}\right) = \exp\left(-\frac{\sqrt{\Vec{p}^2 + m^2} - \mu}{T}\right).
\end{align}
Recall that the energy, pressure, and number density appear in constitutive relations
\begin{align}
    &T_{eq}^{\alpha \beta} = \varepsilon_{eq} u^\alpha u^\beta - P_{eq} \Delta^{\alpha \beta},\no\\
    &j_{eq}^\mu = n_{eq} u^\mu.
\end{align}
where \(\Delta^{\alpha \beta} = \eta^{\alpha \beta} - u^\alpha u^\beta\). \footnote{We choose the notation convention that $\eta_{\mu \nu} = \mbox{diag}(1, -1, -1, -1)$ and $u_\mu u^\mu = 1$.}
Combining the latter equation with definitions given in Eq. \eqref{eq: def-T- J} leads us to \cite{Florkowski:2014sfa}
\begin{align}\label{eq:thermo}
  & \varepsilon_{eq} =   T_{eq}^{\alpha \beta} u_\alpha u_\beta = \int \, \frac{d^3p}{(2\pi)^3} \, \frac{(p \cdot u)^2}{p^0} f_{eq} = \frac{m^2 T}{2 \pi^2} \left( m K_1(\frac{m}{T}) + 3 T K_2(\frac{m}{T})\right),\no\\
  & P_{eq} = - \frac{1}{3} \Delta_{\alpha \beta} T_{eq}^{\alpha \beta} = \frac{1}{3} \int \, \frac{d^3p}{(2\pi)^3} \, \frac{p^2}{p^0} f_{eq} = \frac{m^2 T^2}{2 \pi^2} K_2(\frac{m}{T}),\no\\
  & n_{eq} = u_\mu j_{eq}^\mu = \int \, \frac{d^3p}{(2\pi)^3} \, \frac{p \cdot u}{p^0} f_{eq} = \frac{m^2 T}{2 \pi^2} K_2(\frac{m}{T}),\no\\
  &S_{eq} = \frac{\varepsilon_{eq} + P_{eq}}{T} = \frac{m^2}{2 \pi^2} \left( m K_1(\frac{m}{T}) + 4 T K_2(\frac{m}{T})\right) = \frac{\partial P_{eq}(T)}{\partial T} \bigg|_{T},
\end{align}
where \(K_n(x)\) are modified Bessel functions of the second kind. More detailed explanations about these functions are given in Appendix \ref{sec-integrals}. From these relations, we can find the trace anomaly and speed of sound
\begin{align}\label{eq:trace-cs2}
    & \Theta \equiv \frac{(T^\mu_{\mu})_{eq}}{T^4} = \frac{\varepsilon_{eq} - 3 P_{eq}}{T^4} = \frac{x^3 K_1(x)}{2\pi^2},\no\\
    & c_s^2 = \frac{\partial P_{eq}}{\partial \varepsilon_{eq}} = \frac{ K_3(x)}{x K_2(x) + 3 K_3(x)}.
\end{align}
In Fig. \ref{fig: plot-cs2-trace}, we present the trace anomaly (left panel) and the speed of sound (right panel) as functions of \( x \). At \( x = 0 \) (the massless limit), we find \( \Theta = 0 \) and \( c_s^2 = 1/3 \). In contrast, as \( x \to \infty \) (the heavy quasiparticle limit), both \( c_s^2 \) and \( \Theta \) approach zero. This indicates that conformal symmetry is restored in both the massless and infinitely massive limits. However, the trace anomaly reaches its maximum value, \( \Theta_{\text{Max}} \sim 0.058 \), at \( x = 2.386 \), demonstrating a significant departure from scale invariance at this point.
\section{Charge Diffusion Correlators}
In this section, we calculate the charge correlation functions for a system of massive particles characterized by a constant temperature \( T = T_0 \) and a static fluid velocity \( u^\mu = (1, 0, 0, 0) \). We examine the system's linear response to a small external perturbation \(\delta A_\mu\), while explicitly excluding metric perturbations from our consideration. The correlation functions are obtained by solving Eq. \eqref{eq:eq-BE} within the small-fluctuation approximation and substituting solutions into Eq. \eqref{eq: def-T- J}. Notably, the chemical potential enters only through the fluctuation terms, with the static background solutions maintaining \(\mu_0 = 0\). For first-order fluctuations, the Boltzmann equation in Eq. \eqref{eq:eq-BE} reduces to
\begin{align}
    \left( \partial_t + \Vec{v} \cdot \Vec{\partial}\right) \delta f - \frac{f_{eq}}{T_0} \, \Vec{v} \cdot \Vec{E} = - \frac{1}{\tau_{eq}} \left( \delta f - \delta f_{eq}\right),
\end{align}
where \(\Vec{v}(\Vec{p}) = \frac{\Vec{p}}{p_0}\) is the particle's velocity, and \(\Vec{E} = \Vec{\partial} A_0 - \partial_t \Vec{A} \) is the perturbing electric field. According to Eq. \eqref{eq: equ-MB}
\begin{align}
    \delta f_{eq} = \delta \mu \frac{f_{eq}}{T_0}.
\end{align}
By introducing the Fourier transform
\begin{align}
    \delta f (t, \Vec{x}) = \int d^4 k \, \exp{(- i \omega t + i \Vec{k} \cdot \Vec{x})} \delta f(\omega, \Vec{k}),
\end{align}
we obtain the response one-particle distribution function
\begin{align}
    \delta f(\omega, \Vec{k}, \Vec{p}) = \frac{f_{eq}}{T_0} \, \frac{\delta \mu + \tau_{eq} \Vec{v} \cdot \Vec{E}}{1 + \tau_{eq} (- i \omega + i \Vec{k} \cdot \Vec{v}(\Vec{p}))}. 
\end{align}
This fluctuating function gives rise to the following density and current fluctuations
\begin{align}\label{eq:def-corr}
    &\delta n(\omega, \Vec{k}) = \frac{\delta J^0}{q} = \int \, \frac{d^3p}{(2\pi)^3} \, \delta f(\omega, \Vec{k}, \Vec{p}) = \int \frac{dp}{2\pi^2 T_0}\, p^2 f_{eq}(p) \, \int \frac{d\Omega}{4\pi} \frac{\delta \mu + \tau_{eq} \Vec{v}(\Vec{p}) \cdot \Vec{E}}{1 + \tau_{eq} (- i \omega + i \Vec{k} \cdot \Vec{v}(\Vec{p}))}, \no\\
    &\delta j^i(\omega, \Vec{k}) = \frac{\delta J^i}{q} = \int \, \frac{d^3p}{(2\pi)^3} \, \frac{p^i}{p_0}\, \delta f(\omega, \Vec{k}, \Vec{p}) = \int \frac{dp}{2\pi^2 T_0}\, p^2 f_{eq}(p) \, \int \frac{d\Omega}{4\pi} \, v^i(\Vec{p})\, \frac{\delta \mu + \tau_{eq} \Vec{v}(\Vec{p}) \cdot \Vec{E}}{1 + \tau_{eq} (- i \omega + i \Vec{k} \cdot \Vec{v}(\Vec{p}))},
\end{align}
where \(p \equiv \vert \Vec{p} \vert\). To simplify these equations, we use \(\delta n = \chi \, \delta \mu\), where \(\chi\) is the number susceptibility defined as
\begin{align}
   \chi(m, T_0) = \int \frac{dp}{2\pi^2 T_0}\, p^2 f_{eq}(p) = \frac{T_0^2}{2\pi^2} \, x^2 K_2(x),
\end{align}
The susceptibility function behaves at extreme limits as follows
\begin{align}
    & \chi(m, T_0)\bigg|_{m\ll T_0} = \frac{T_0^2}{2\pi^2} \,\left(2 - \frac{1}{2} x^2 + \mathcal{O}\left(x^4\right)\right), \no\\
    & \chi(m, T_0)\bigg|_{m\gg T_0} = \frac{T_0^2}{2\pi^2} \, e^{-x} \left( x^{\frac{3}{2}} \sqrt{\frac{\pi}{2}} + \mathcal{O}\left(x^{\frac{1}{2}}\right)\right).
\end{align}
The susceptibility definition turns Eq. \eqref{eq:def-corr} into the following relations
\begin{align}
    &\delta n(\omega, \Vec{k}) = \frac{\int \frac{dp}{2\pi^2 T_0}\, p^2 f_{eq}(p) \, \int \frac{d\Omega}{4\pi} \frac{\tau_{eq} \Vec{v}(\Vec{p}) \cdot \Vec{E}}{1 + \tau_{eq} (- i \omega + i \Vec{k} \cdot \Vec{v}(\Vec{p}))}}{1 - \frac{1}{2 i k \tau_{eq} \chi} \int \frac{dp}{2\pi^2 T_0}\, p^2 f_{eq}(p) \, \frac{1}{v} \, \ln{\left( \frac{\omega - k \, v + \frac{i}{\tau_{eq}}}{\omega + k \, v + \frac{i}{\tau_{eq}}}\right)}}, \no\\
    &\delta j^i(\omega, \Vec{k}) = \int \frac{dp}{2\pi^2 T_0}\, p^2 f_{eq}(p) \, \int \frac{d\Omega}{4\pi} \, \hat{p}^i\, \frac{\frac{\delta n(\omega, \Vec{k})}{\chi} + \tau_{eq} \Vec{v}(\Vec{p}) \cdot \Vec{E}}{1 + \tau_{eq} (- i \omega + i \Vec{k} \cdot \Vec{v}(\Vec{p}))}.
\end{align}
To obtain these results, we have used
\begin{align}
    \int \frac{d\Omega}{4\pi} \frac{1}{1 + \tau_{eq} (- i \omega + i \Vec{k} \cdot \Vec{v}(\Vec{p}))} = \frac{1}{2 i k \tau_{eq} v} \ln{\left( \frac{\omega - k \, v + \frac{i}{\tau_{eq}}}{\omega + k \, v + \frac{i}{\tau_{eq}}}\right)},
\end{align}
where \(v = \frac{p}{\sqrt{p^2 + m^2}}\) and \(k = \vert \Vec{k}\vert\). This demonstrates that the massive case leads to different correlation functions compared to the massless case \cite{Romatschke:2015gic}. From definitions of correlation functions given in Eq. \eqref{eq: def-var}, we obtain the retarded current-current two-point function as follows
\begin{align}\label{eq: corr-charge}
    &G_{0 0}(\omega, k) = - \frac{\delta n}{\delta A_0}\bigg|_{\delta A =  0} =  \, \frac{\int \frac{dp}{2\pi^2 T_0}\, p^2 f_{eq}(p) \, \mathcal{I}_1(v, \omega, k)}{1 - \frac{1}{2 i k \tau_{eq} \chi} \int \frac{dp}{2\pi^2 T_0}\, p^2 f_{eq}(p) \, \frac{1}{v} \, \ln{\left( \frac{\omega - k \, v + \frac{i}{\tau_{eq}}}{\omega + k \, v + \frac{i}{\tau_{eq}}}\right)}},\no\\
    &G_{0 3}(\omega, k) =  - \frac{\delta n}{\delta A^3}\bigg|_{\delta A =  0} =  \frac{\omega}{k}\,  \frac{\int \frac{dp}{2\pi^2 T_0}\, p^2 f_{eq}(p) \,\mathcal{I}_1(v, \omega, k)}{1 - \frac{1}{2 i k \tau_{eq} \chi} \int \frac{dp}{2\pi^2 T_0}\, p^2 f_{eq}(p) \, \frac{1}{v} \, \ln{\left( \frac{\omega - k \, v + \frac{i}{\tau_{eq}}}{\omega + k \, v + \frac{i}{\tau_{eq}}}\right)}},\no\\
    &G_{3 3}(\omega, k) =   - \frac{\delta j_3}{\delta A^3}\bigg|_{\delta A =  0} = \frac{G_{0 3}(\omega, \Vec{k})}{\chi}\, \int \frac{dp}{2\pi^2 T_0}\, p^2 f_{eq}(p) \, \mathcal{I}_2(v, \omega, k) + \int \frac{dp}{2\pi^2 T_0}\, p^2 f_{eq}(p) \, \mathcal{I}_3(v, \omega, k),\no\\
    &G_{1 1}(\omega, k) =  G_{2 2}(\omega, k) = - \frac{\delta j_1}{\delta A^1}\bigg|_{\delta A  = 0} = \int \frac{dp}{2\pi^2 T_0}\, p^2 f_{eq}(p) \, \mathcal{I}_4(v, \omega, k).
\end{align}
The unknown \(\left( \mathcal{I}_i(v, \omega, k), \, i= 1, \cdots, 4 \right)\) functions are defined in the Appendix. \ref{sec-definitions}. There is no way to calculate these integrals directly. The only way is to use some approximations. We follow the expansion strategy in small and large \(x = m/T_0\). To do this, we need to know the expansion in ``\(x\)'' for the following function 
\begin{align}\label{eq-smallx}
    &\frac{1}{v} \ln{\left( \frac{\omega - k \, v + \frac{i}{\tau_{eq}}}{\omega + k \, v + \frac{i}{\tau_{eq}}}\right)}\bigg|_{x \ll 1} = L(\omega, k) + \frac{x^2}{2 \Tilde{p}^2} \left( L(\omega, k) - \frac{2 i k \tau_{eq} (1 - i \tau_{eq} \omega)}{(1 - i \tau_{eq} \omega)^2 + k^2 \tau_{eq}^2}\right)\no\\
   & \hspace{4.25cm} - \frac{x^4}{8 \Tilde{p}^4} \left( L(\omega, k) + \frac{2 i k \tau_{eq} (1 - i \tau_{eq} \omega) \left( k^2 \tau_{eq}^2 - (1 - i \tau_{eq} \omega)^2 \right)}{\left((1 - i \tau_{eq} \omega)^2 + k^2 \tau_{eq}^2\right)^2}\right) + \mathcal{O}(\frac{x^6}{ \Tilde{p}^6}),\no\\
   & \frac{1}{v} \ln{\left( \frac{\omega - k \, v + \frac{i}{\tau_{eq}}}{\omega + k \, v + \frac{i}{\tau_{eq}}}\right)}\bigg|_{x \gg 1} = \frac{2 i k \tau_{eq}}{1 - i \tau_{eq} \omega} - \frac{2 i \Tilde{p}^2 k^3 \tau_{eq}^3}{3 x^2 (1 - i \tau_{eq} \omega)^3} + \frac{2 i \Tilde{p}^4 k^3 \tau_{eq}^3 \left( 3 k^2 \tau_{eq}^2 + 5 (1 - i \tau_{eq} \omega)^2\right)}{15 x^4 (1 - i \tau_{eq} \omega)^5} + \mathcal{O}(\frac{\Tilde{p}^6}{x^6}),
\end{align}
where \(\Tilde{p} = \frac{p}{T_0}\) and \(L(\omega, k) \equiv \ln{\left( \frac{\omega - k  + \frac{i}{\tau_{eq}}}{\omega + k  + \frac{i}{\tau_{eq}}}\right)}\). Also, it is necessary to know
\begin{align}\label{eq:expand-v}
   & v \ln{\left( \frac{\omega - k \, v + \frac{i}{\tau_{eq}}}{\omega + k \, v + \frac{i}{\tau_{eq}}}\right)}\bigg|_{x \ll 1} = L(\omega, k) - \frac{x^2}{2 \Tilde{p}^2} \left( L(\omega, k) + \frac{2 i k \tau_{eq} (1 - i \tau_{eq} \omega)}{(1 - i \tau_{eq} \omega)^2 + k^2 \tau_{eq}^2}\right)\no\\
   & \hspace{4cm} + \frac{x^4}{8 \Tilde{p}^4} \left( 3 L(\omega, k) + \frac{2 i k \tau_{eq} (1 - i \tau_{eq} \omega) \left( 3 k^2 \tau_{eq}^2 +5 (1 - i \tau_{eq} \omega)^2 \right)}{\left((1 - i \tau_{eq} \omega)^2 + k^2 \tau_{eq}^2\right)^2}\right) + \mathcal{O}(\frac{x^6}{ \Tilde{p}^6}),\no\\
   & v \ln{\left( \frac{\omega - k \, v + \frac{i}{\tau_{eq}}}{\omega + k \, v + \frac{i}{\tau_{eq}}}\right)}\bigg|_{x \gg 1} =  \frac{2 i \Tilde{p}^2 k \tau_{eq}}{x^2 (1 - i \tau_{eq} \omega)} - \frac{2 i \Tilde{p}^4 k \tau_{eq} \left(  k^2 \tau_{eq}^2 + 3 (1 - i \tau_{eq} \omega)^2\right)}{3 x^4 (1 - i \tau_{eq} \omega)^3} + \mathcal{O}(\frac{\Tilde{p}^6}{x^6}).
\end{align}
Doing these expansions makes it feasible to perform `` $p$'' integrals. In the limit of small $x$ according to Eq. \eqref{eq-smallx} and given integrals in Appendix \ref{sec-integrals}, we obtain
\begin{align}
   &\mathcal{S}^{(s)}(x; \omega, \Vec{k}) \equiv \int \frac{dp}{2\pi^2 T_0}\, p^2 f_{eq}(p) \, \frac{1}{v} \, \ln{\left( \frac{\omega - k \, v + \frac{i}{\tau_{eq}}}{\omega + k \, v + \frac{i}{\tau_{eq}}}\right)}\bigg|_{x \ll 1} \no\\
   &= \chi(x) \bigg( L(\omega, k) + \frac{x K_1(x)}{2 K_2(x)}  \left( L(\omega, k) - \frac{2 i k \tau_{eq} (1 - i \tau_{eq} \omega)}{(1 - i \tau_{eq} \omega)^2 + k^2 \tau_{eq}^2}\right) \no\\
   &+ \frac{x^2 K_0(x)}{8 K_2(x)} \left( L(\omega, k) + \frac{2 i k \tau_{eq} (1 - i \tau_{eq} \omega) \left( k^2 \tau_{eq}^2 - (1 - i \tau_{eq} \omega)^2 \right)}{\left((1 - i \tau_{eq} \omega)^2 + k^2 \tau_{eq}^2\right)^2}\right) + \cdots\bigg),\no\\
  &\mathcal{S}^{(l)}(x; \omega, \Vec{k}) \equiv \int \frac{dp}{2\pi^2 T_0}\, p^2 f_{eq}(p) \, \frac{1}{v} \, \ln{\left( \frac{\omega - k \, v + \frac{i}{\tau_{eq}}}{\omega + k \, v + \frac{i}{\tau_{eq}}}\right)}\bigg|_{x \gg 1} \no\\
  &=  \chi(x) \bigg( \frac{2 i k \tau_{eq}  }{1 - i \tau_{eq} \omega} - \frac{2 i k^3 \tau_{eq}^3 K_3(x)}{ x  K_2(x) (1 - i \tau_{eq} \omega)^3} + \frac{2 i k^3 \tau_{eq}^3 \left( 3 k^2 \tau_{eq}^2 + 5 (1 - i \tau_{eq} \omega)^2\right) K_4(x)}{ x^2  K_2(x) \, (1 - i \tau_{eq} \omega)^5}\bigg).
\end{align}
The superscript $s(l)$ refers to the small (large) limits of $x$. Similarly, integrals of expressions in Eq. \eqref{eq:expand-v} can be performed
\begin{align}
   &\mathcal{T}^{(s)}(x; \omega, \Vec{k}) \equiv \int \frac{dp}{2\pi^2 T_0}\, p^2 f_{eq}(p) \, v \, \ln{\left( \frac{\omega - k \, v + \frac{i}{\tau_{eq}}}{\omega + k \, v + \frac{i}{\tau_{eq}}}\right)}\bigg|_{x \ll 1} \no\\
   &= \chi(x) \bigg( L(\omega, k) - \frac{x K_1(x)}{2 K_2(x)}  \left( L(\omega, k) + \frac{2 i k \tau_{eq} (1 - i \tau_{eq} \omega)}{(1 - i \tau_{eq} \omega)^2 + k^2 \tau_{eq}^2}\right) \no\\
   &- \frac{x^2 K_0(x)}{8 K_2(x)} \left( 3 L(\omega, k) + \frac{2 i k \tau_{eq} (1 - i \tau_{eq} \omega) \left( 4 k^2 \tau_{eq}^2 +5 (1 - i \tau_{eq} \omega)^2 \right)}{\left((1 - i \tau_{eq} \omega)^2 + k^2 \tau_{eq}^2\right)^2}\right) + \cdots\bigg),\no\\
   & \mathcal{T}^{(l)}(x; \omega, \Vec{k}) \equiv \int \frac{dp}{2\pi^2 T_0}\, p^2 f_{eq}(p) \, v \, \ln{\left( \frac{\omega - k \, v + \frac{i}{\tau_{eq}}}{\omega + k \, v + \frac{i}{\tau_{eq}}}\right)}\bigg|_{x \gg 1} \no\\
   &= \chi(x) \bigg( \frac{6 i k \tau_{eq} K_3(x) }{x K_2(x) (1 - i \tau_{eq} \omega)} - \frac{10 i k \tau_{eq} \left(  k^2 \tau_{eq}^2 + 3 (1 - i \tau_{eq} \omega)^2\right) K_4(x)}{x^2 K_2(x) (1 - i \tau_{eq} \omega)^3}\bigg).
\end{align}
Eventually, we derive the expressions for retarded charge correlators given in Eq. \eqref{eq: corr-charge} as follows
\begin{subequations}
\begin{align}\label{eq:final-corr-Rcharge}
    &G^{(a)}_{0 0}(x; \omega, \Vec{k}) = \chi(x) \frac{ 1 + \frac{i \omega \tau_{eq} - 1}{2 i \tau_{eq} k \chi(x)} \mathcal{S}^{(a)}(x; \omega, \Vec{k})}{1 - \frac{\mathcal{S}^{(a)}(x; \omega, \Vec{k})}{2 i k \tau_{eq} \chi(x)}},\\
    & G^{(a)}_{0 3}(x; \omega, \Vec{k}) = \frac{\omega}{k} G^{(a)}_{0 0}(x; \omega, \Vec{k}),\\
    & G^{(a)}_{3 3}(x; \omega, \Vec{k}) = \frac{\omega^2}{k^2} G^{(a)}_{0 0}(x; \omega, \Vec{k}),\\
    & G^{(a)}_{1 1}(x; \omega, \Vec{k}) = G^{(a)}_{2 2}(x; \omega, \Vec{k})= - \frac{i \tau_{eq} \omega \chi(x)}{2} \left( \frac{1 - i \tau_{eq} \omega}{(\tau_{eq}\, k )^2} + \frac{(1 - i \tau_{eq} \omega)^2 \mathcal{S}^{(a)}(x; \omega, \Vec{k})}{2  (i \tau_{eq}\, k )^3 \chi(x)}  + \frac{i \mathcal{T}^{(a)}(x; \omega, \Vec{k})}{2 k \tau_{eq} \chi(x)} \right).
\end{align}
\end{subequations}
where $a= (s, l)$. For a massless system of particles, i.e. $x \to 0$, we obtain the previous results \cite{Romatschke:2015gic}.
\subsection{Hydrodynamic limit}
\begin{figure}
    \centering
    \includegraphics[width=0.475\textwidth, valign=t]{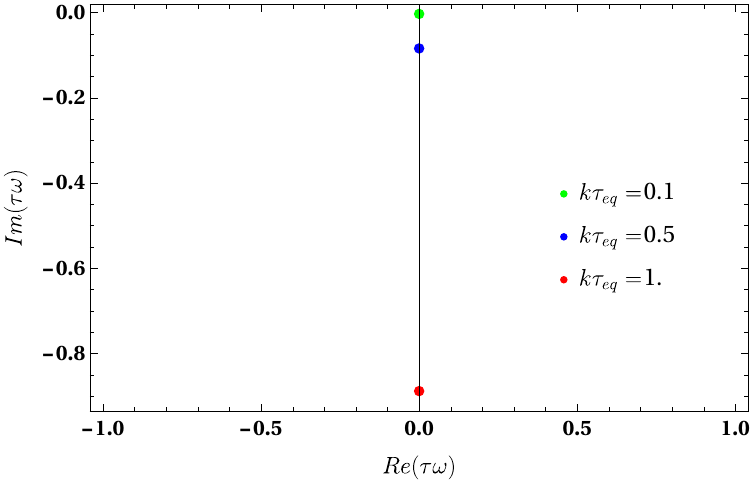}
\hspace{0.3cm}
\includegraphics[width=0.488\textwidth,valign=t]{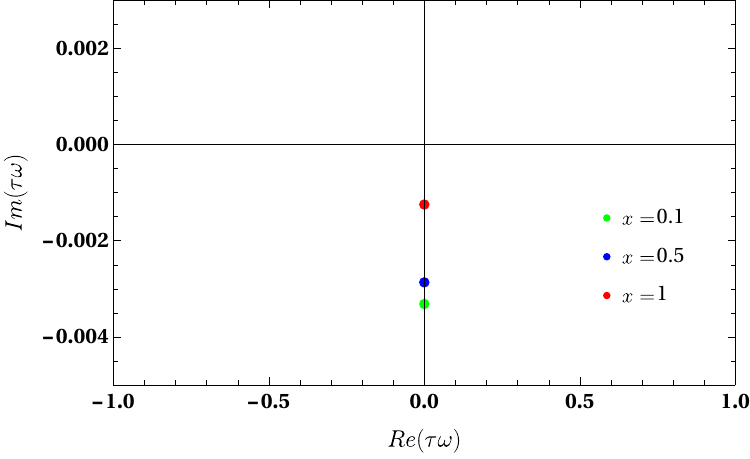}
\caption{Diffusive hydro modes for different cases. Left: pole structure of $G_{0 0}$ at $x = 0.1$ for different momenta. Right: the same structure at $k \tau = 0.1$ for different mass-to-temperature ratios. Different colors show different values of ``$x (k \tau)$'' in the right (left) panels.}
    \label{fig: plot-x-kt}
\end{figure}

At small momenta and frequencies, hydrodynamic modes are identified as poles in the retarded correlator. For small $x$, the expansion of correlator in Eq. \eqref{eq:final-corr-Rcharge} reads as follows
\begin{align}\label{eq:G00-l}
    &G^{(s)}_{0 0}(x; \omega, \Vec{k}) = \chi(x) \frac{\mathcal{N}^{0 (s)}(x; \omega, \Vec{k})}{\mathcal{D}^{0 (s)}(x; \omega, \Vec{k})},
\end{align}
where
\begin{align}\label{eq:G00-detail}
  & \mathcal{N}^{0 (s)}(x; \omega, \Vec{k}) = \frac{i (1 - i) (1 - i \tau_{eq} \omega)}{\tau_{eq}\omega } - \frac{i k^2 \tau_{eq} N(x)}{15 \omega (1 - i \tau_{eq} \omega) K_2(x)\left(x^2 K_0(x) + x K_1(x) - K_2(x)\right)},\no\\
  & \mathcal{D}^{0 (s)}(x; \omega, \Vec{k}) = 1 - \frac{i k^2 \tau_{eq} D(x)}{15 \omega (1 - i \tau_{eq} \omega)^2 K_2(x)\left(x^2 K_0(x) + x K_1(x) - K_2(x)\right)},
  \end{align}
 in which
  \begin{align}
  & N(x) = 5 i x^2 \left(x K_0(x) + K_1(x)\right)^2 - x K_2(x) (1 - i) \left( x K_0(x) (22 + 5 i) + K_1(x)(13 + 5 i)\right) + (9- 4 i) K_2^2(x),\no\\
  & D(x) = 5 x^2 \left(x K_0(x) + K_1(x)\right)^2 - x K_2(x) \left( x K_0(x) (10 - 27 i \tau_{eq} \omega) + 2 K_1(x)(5 - 9  i \tau_{eq} \omega)\right) + (5- 9 i \tau_{eq} \omega) K_2^2(x).
\end{align}
The lowest lying modes are solutions of spectral curves $\mathcal{D}^{(l)}(x; \omega, \Vec{k}) = 0$ at small momenta which are obtained as follows
\begin{align}\label{eq:hydro-Rcharge-small-x}
    \omega^{(nh)}_{(1, 2)}\bigg\vert_{x \ll 1} &= - \frac{i}{\tau_{eq}} \pm \frac{2 k}{\sqrt{15}} \left(1 - \frac{x^2}{4} + \cdots \right)+ \frac{i k^2 \tau_{eq}}{6} \left( 1 - \frac{x^2}{2} + \cdots \right) + \mathcal{O}(k^3),\no\\
    \omega^{(h)}\bigg\vert_{x \ll 1} &= - \frac{i k^2 \tau_{eq}}{3} \left( 1 - \frac{x^2}{2}  + \cdots \right) + \mathcal{O}(k^4),
\end{align}
where ``$nh$'' and ``$h$'' refer to non-hydro and hydro modes, respectively. It can be easily verified that the $x \to 0$ limit of the latter relations gives the massless results \cite{Romatschke:2015gic}. On the other hand, the expansion of retarded correlator in the large $x$ limit yields
\begin{align}
    &G^{(l)}_{0 0}(x; \omega, \Vec{k})\bigg\vert_{x \gg 1} = \chi(x) \frac{\mathcal{N}^{0 (l)}(x; \omega, \Vec{k})}{\mathcal{D}^{0 (l)}(x; \omega, \Vec{k})},
\end{align}
where
\begin{align}\label{eq:G00-detail-hx}
  & \mathcal{N}^{0 (l)}(x; \omega, \Vec{k}) = k^2 \tau_{eq}^2 (1 - i \tau_{eq} \omega) \left( x K_3(x) - 5 K_4(x)\right)^2,\\
  & \mathcal{D}^{0 (l)}(x; \omega, \Vec{k}) = k^2 \tau_{eq}^2 \left( x K_3(x) - 5 K_4(x)\right)^2 - i \tau_{eq} \omega x^2 K_2(x) \bigg( x K_3(x) (1 - i \tau_{eq} \omega) + (3 k^2 \tau_{eq}^2 - 5 (1 - i \tau_{eq} \omega)^2) K_4(x)\bigg).\no
  \end{align}
In the large $x$ limit, the lowest modes are given as follows
\begin{align}\label{eq:hydro-Rcharge-large-x}
    \omega^{(nh)}_{(1, 2)}\bigg\vert_{x \gg 1} &= - \frac{i}{\tau_{eq}} \pm k \sqrt{\frac{2}{x}}+ \frac{i k^2 \tau_{eq}}{2 x} + \mathcal{O}(k^3),\no\\
    \omega^{(h)}\bigg\vert_{x \gg 1} &= - \frac{i k^2 \tau_{eq}}{x} + \mathcal{O}(k^4),
\end{align}
This implies that for extremely massive quasiparticles, the hydro and non-hydro modes are stable. 
In Fig. \ref{fig: plot-x-kt} we plot the diffusive hydro modes for different cases. The left figure indicates the modes at $x = 0.1$ for different $k \tau_{eq}$ values. The right figure shows the same results for $k \tau_{eq} = 0.1$ for different $x$ values. It can be seen that the mass moves the poles into the unstable zone. This is because mass changes the diffusion constant, thereby influencing the location of hydro poles. 



Retarded correlations enable us to obtain transport coefficients from the Kubo formulas \cite{Kovtun:2012rj}. In the charge correlation functions, the diffusion constant $D$ and diffusion relaxation time $\tau_D$ can be derived as \cite{Romatschke:2015gic, Drude:1900dr}
\begin{align}
    \lim_{\omega\rightarrow 0}\lim_{k\rightarrow 0}\frac{\omega}{k^2}{\rm Im}\,G_{00}(\omega, k)&=  \chi D\,\no\\
\lim_{\omega\rightarrow 0}\lim_{k\rightarrow 0}\frac{1}{k^2}{\rm Re}\,G_{00}(\omega, k)&=  \chi D\tau_D.
\end{align}
For small $x$, using the expressions given in Eqs. \eqref{eq:G00-l} and \eqref{eq:G00-detail} will readily obtain
\begin{align}
    &D(x)\bigg\vert_{x \ll 1} = \frac{\tau_{eq}}{3} \left(1 - \frac{x^2}{2} + \cdots \right),\no\\
    &\tau_D\bigg\vert_{x \ll 1}  = \tau_{eq}.
\end{align}
For large $x$, the corresponding transport coefficients are 
\begin{align}
    &D(x)\bigg\vert_{x \gg 1} = \tau_{eq}\left(\frac{1}{x} + \mathcal{O}(\frac{1}{x^2})\right),\no\\
    &\tau_D\bigg\vert_{x \gg 1}  = \tau_{eq}.
\end{align}
\section{Energy-Momentum Correlators}
To calculate the retarded correlation functions for the energy-momentum tensor components, we maintain a fixed chemical potential $\mu = 0$ while allowing temperature and velocity to vary as 
\begin{align}
    T(t, \Vec{x}) = T_0 + \delta T(t, \Vec{x}), \qquad u^\mu(t, \Vec{x}) = (1, 0, 0, 0) + \delta u^\mu(t, \Vec{x}).
\end{align}
The system is analyzed in the presence of small fluctuating metric perturbations
\begin{align}
    g_{\mu \nu} = \eta_{\mu \nu} + \delta g_{\mu \nu}.
\end{align}
Taking these assumptions makes only the gravity force present, as given in Eqs. \eqref{eq: term-F} and  \eqref{eq: term-F-em-g}. This will lead us to the following Boltzmann equation
\footnote{Due to the relation \(\Gamma^\alpha_{\mu \nu} p_\alpha p^\mu p^\nu = p^\mu \partial_\mu p^2 =  0\) for on-shell particles, we may use \(\Gamma_{0 \mu \nu} p_\alpha p^\mu p^0 = \Gamma_{i \mu \nu} p_i p^\mu p^\nu \).}
\begin{align}\label{eq: BE-EMT}
    \left( \partial_t + \Vec{v} \cdot \Vec{\partial}\right) \delta f - \Gamma^0_{\alpha \beta} \frac{p^\alpha p^\beta}{p^0} \frac{\partial}{\partial p^0}f_0 = - \frac{1}{\tau_{eq}} \left( \delta f - \delta f_{eq}\right),
\end{align}
where the fluctuating equilibrium function is given by
\begin{align}
    \delta f_{eq} = \frac{f_{eq} p^0}{T_0} \left( \Vec{v} \cdot \delta \Vec{u} + \frac{\delta T}{T_0}\right).
\end{align}
Solving Eq. \eqref{eq: BE-EMT} in the momentum space yields the following response function
\begin{align}
    \delta f (\omega, \Vec{k}, \Vec{p}) = \frac{f_{eq} p^0}{T_0} \, \frac{\Vec{v} \cdot \delta \Vec{u} + \frac{\delta T}{T_0} - \tau_{eq} \Gamma^0_{\mu \nu} v^\mu v^\nu}{ 1 + \tau_{eq} (- i \omega + i \Vec{k} \cdot \Vec{v})}.
\end{align}
Accordingly, the energy-momentum tensor components will be given as
\begin{align}
    \delta T^{\alpha \beta}(\omega, \Vec{k}) = \int \, \frac{d^3p}{(2\pi)^3} \, \frac{p^\alpha p^\beta}{p^0} \delta f(\omega, \Vec{k}, \Vec{p}) = \int \frac{dp}{2\pi^2 T_0}\, p^2 \, p_0^2 \, f_{eq}(p) \, \int \frac{d\Omega}{4\pi} v^\alpha(\Vec{p}) \, v^\beta(\Vec{p}) \, \frac{\Vec{v} \cdot \delta \Vec{u} + \frac{\delta T}{T_0} - \tau_{eq} \Gamma^0_{\mu \nu} v^\mu v^\nu}{1 + \tau_{eq} (- i \omega + i \Vec{k} \cdot \Vec{v}(\Vec{p}))},
\end{align}
where $v^\alpha(\Vec{p}) = (1, \Vec{v}(\Vec{p}))$. It is worthwhile to mention that $\delta u^i = \frac{\delta T^{0 i}}{\varepsilon_{eq} + P_{eq}}$ and $\frac{\delta T}{T_0} = \frac{\delta T^{0 0} c_s^2}{\varepsilon_{eq} + P_{eq}}$ due to $c_v = \frac{\partial \varepsilon_{eq}}{\partial T} = \frac{S_{eq}}{c_s^2}$, and $\delta \varepsilon = \delta T^{0 0}$. Moreover, the following relation is satisfied due to the integrals given in Appendix \ref{sec-integrals} and the thermodynamic quantities provided in Eqs. \eqref{eq:thermo} and \eqref{eq:trace-cs2}
\begin{align}\label{eq: somes}
    & \int \, \frac{dp}{2 \pi^2 T_0} p^2 \,p_0^2\, f_{eq} (p) = \frac{\varepsilon_{eq} + P_{eq}}{c_s^2} = \frac{T_0^4 x^3}{2\pi^2} \bigg( x K_2(x) + 3 K_3(x)\bigg),\no\\
    &\int \, \frac{dp}{2 \pi^2 T_0} p^2 \,p_0^2\, v^2\, f_{eq} (p) = 3 \left(\varepsilon_{eq} + P_{eq}\right) = \frac{3 T_0^4}{2 \pi^2} x^3 K_3(x),
\end{align}
where \(v^2 = p^2/p_0^2\). Notably, the latter relations are also satisfied for the massless case as well \cite{Romatschke:2015gic}. After some algebraic manipulations, we get the following relations for energy-momentum tensor perturbations
\begin{align}\label{eq:main-Tmunu}
    & \delta T^{0 0} = \frac{c}{a c - b^2 c_s^2} \mathcal{E}_0 + \frac{b}{a c - b^2 c_s^2} \mathcal{E}_3, \no\\
    & \delta T^{0 3} = \frac{b \, c_s^2}{a c - b^2 c_s^2} \mathcal{E}_0 + \frac{a}{a c - b^2 c_s^2} \mathcal{E}_3, \\
    & \delta T^{0 i} = -\frac{ \tau_{eq}}{d} \int \, \frac{dp}{2 \pi^2 T_0} p^2 \,p_0^2\, f_{eq} (p) \bigg\{ \Gamma^0_{0 i}(\omega, \Vec{k}) \frac{\mathcal{I}_4(v, \omega, k)}{i \tau_{eq} \omega} + \Gamma^0_{3 i}(\omega, \Vec{k}) \mathcal{I}_6(v, \omega, k)\bigg\}, \,\, i = 1, 2, \no\\
    & \delta T^{3 3} = \frac{\delta T^{0 0} c_s^2 (1 - i \tau_{eq} \omega)}{k^2 \tau_{eq}^2(\varepsilon_{eq} + P_{eq})} \int \, \frac{dp}{2 \pi^2 T_0} p^2 \,p_0^2\, f_{eq} (p)  \mathcal{I}_1(v, \omega, k) + \frac{\delta T^{0 3}}{\varepsilon_{eq} + P_{eq}} \int \, \frac{dp}{2 \pi^2 T_0} p^2 \,p_0^2\, f_{eq} (p) \mathcal{I}_5(v, \omega, k)\no\\
    &\hspace{0.75cm} - \tau_{eq} \int \, \frac{dp}{2 \pi^2 T_0} p^2 \,p_0^2\, f_{eq} (p) \bigg\{ \Gamma^0_{0 0}(\omega, \Vec{k}) \frac{1 - i \tau_{eq} \omega}{ k^2 \tau_{eq}^2} \mathcal{I}_1(v, \omega, k) + 2 \, \Gamma^0_{0 3}(\omega, \Vec{k}) \mathcal{I}_5(v, \omega, k) + \Gamma^0_{3 3}(\omega, \Vec{k}) \mathcal{I}_7(v, \omega, k)\no\\
    &\hspace{5cm}+ (\Gamma^0_{1 1}(\omega, \Vec{k}) + \Gamma^0_{2 2}(\omega, \Vec{k})) \mathcal{I}_8(v, \omega, k) \bigg\}, \no\\
    & \delta T^{i i} = \frac{\delta T^{0 0} c_s^2 }{i \tau_{eq} \omega (\varepsilon_{eq} + P_{eq})} \int \, \frac{dp}{2 \pi^2 T_0} p^2 \,p_0^2\, f_{eq} (p)  \mathcal{I}_4(v, \omega, k) + \frac{\delta T^{0 3}}{\varepsilon_{eq} + P_{eq}} \int \, \frac{dp}{2 \pi^2 T_0} p^2 \,p_0^2\, f_{eq} (p) \mathcal{I}_6(v, \omega, k)\no\\
    & \hspace{0.75cm} - \tau_{eq} \int \, \frac{dp}{2 \pi^2 T_0} p^2 \,p_0^2\, f_{eq} (p) \bigg\{ \Gamma^0_{0 0}(\omega, \Vec{k}) \frac{\mathcal{I}_4(v, \omega, k)}{ i \tau_{eq} \omega}  + 2\, \Gamma^0_{0 3}(\omega, \Vec{k}) \mathcal{I}_6(v, \omega, k) +  \Gamma^0_{3 3}(\omega, \Vec{k}) \mathcal{I}_8(v, \omega, k)\no\\
    &\hspace{5cm}+ (2 \Gamma^0_{i i}(\omega, \Vec{k})+ \Gamma^0_{1 1}(\omega, \Vec{k}) + \Gamma^0_{2 2}(\omega, \Vec{k})) \mathcal{I}_9(v, \omega, k) \bigg\}, \,\,\, i = 1, 2, \no\\
    & \delta T^{1 2} = - \, \tau_{eq} \int \, \frac{dp}{2 \pi^2 T_0} p^2 \,p_0^2\, f_{eq} (p) \Gamma^0_{1 2}(\omega, \Vec{k}) \mathcal{I}_9(v, \omega, k),\no\\
    & \delta T^{3 i} = \frac{\delta T^{0 i}}{\varepsilon_{eq} + P_{eq}} \int \, \frac{dp}{2 \pi^2 T_0} p^2 \,p_0^2\, f_{eq} (p) \mathcal{I}_6(v, \omega, k)\no\\
    & \hspace{0.75cm} - \, \tau_{eq} \int \, \frac{dp}{2 \pi^2 T_0} p^2 \,p_0^2\, f_{eq} (p) \bigg\{ \Gamma^0_{0 i}(\omega, \Vec{k}) \mathcal{I}_6(v, \omega, k) + \Gamma^0_{3 i}(\omega, \Vec{k})\mathcal{I}_8(v, \omega, k)\bigg\}, \,\, i = 1, 2.\no
\end{align}
For the sake of brevity, we define the unknown variables and functions in Appendix \ref{sec-definitions}. The following relations hold between ($a, b, c$) parameters
\begin{align}\label{eq:a-b-c}
    &a = \frac{\omega - c_s^2 k b}{\omega + \frac{i}{\tau_{eq}}}, \qquad c = 1 - \frac{\omega + \frac{i}{\tau_{eq}}}{k} b,\no\\
    & a c - b^2 c_s^2 = \frac{\omega - b \left( c_s^2 k + \frac{\omega}{k}(\omega + \frac{i}{\tau_{eq}}) \right)}{\omega + \frac{i}{\tau_{eq}}}.
\end{align}
These equations are derivable from Eq. \eqref{eq:App-B1}, which means that knowing only ``$b$'' is sufficient to determine the spectrum of correlators. Based on the definition of the Christoffel symbol given in Eq. \eqref{eq: term-F-em-g}, the relevant components of the Christoffel symbol in the first order of metric fluctuations take the following form
\begin{align}\label{eq: christofel}
    &\Gamma^0_{0 0}(t, \Vec{x}) = \frac{\eta^{00}}{2}\, \partial_0 \delta g_{0 0}(t, \Vec{x}) \rightarrow \,\, \Gamma^0_{0 0}(\omega, \Vec{k}) = - \frac{i \omega}{2} \delta g_{0 0}, \no\\
    &\Gamma^0_{0 j}(t, \Vec{x}) = \frac{\eta^{00}}{2}\, \partial_j \delta g_{0 0}(t, \Vec{x}) \rightarrow \,\, \Gamma^0_{0 j}(\omega, \Vec{k}) =  \frac{i k_j}{2} \delta g_{0 0}, \,\,\, j = 1, 2, 3,\\
    &\Gamma^0_{m n}(t, \Vec{x}) = \frac{\eta^{00}}{2}\, \left( \partial_m \delta g_{n 0}(t, \Vec{x}) + \partial_n \delta g_{m 0}(t, \Vec{x}) - \partial_0 \delta g_{m n}(t, \Vec{x})\right),\no\\
    &\rightarrow \Gamma^0_{m n}(\omega, \Vec{k}) = \frac{i}{2} \left( k_m \delta g_{n 0} + k_n \delta g_{m 0} + \omega \delta g_{ m n}\right),\,\, m, n = 1, 2, 3. \no
\end{align}
Considering the latter equation together with the correlation definition given in Eq. \eqref{eq: def-var}, we obtain the following results for two-point correlations of energy-momentum tensor components
\begin{align}\label{eq:corr-TT-p1}
    &G^{0 0, 0 0}(\omega, k) = - \frac{\varepsilon_{eq} + P_{eq}}{c_s^2} \bigg( 1 + \frac{i \, b \, c_s^2 \, k \, \tau_{eq}}{a c - b^2 c_s^2}\bigg),\no\\
    &G^{0 0, 0 3}(\omega, k) = \frac{\omega}{k}\left(G^{0 0, 0 0}(\omega, k) + \frac{\varepsilon_{eq} + P_{eq}}{c_s^2}\right),\no\\
    &G^{0 0, 3 3}(\omega, k) = \frac{\omega}{k}G^{0 0, 0 3}(\omega, k),\no\\
    & G^{0 0, 1 1}(\omega, k) = G^{0 0, 2 2}(\omega, k) \no\\
    &= \frac{ \varepsilon_{eq} + P_{eq}}{a c - b^2 c_s^2} \frac{\omega}{2 k} \, \bigg(b \left(i \tau_{eq} \omega - 3\right) - \frac{1}{2 (\varepsilon_{eq} + P_{eq})} \, \int \, \frac{dp}{2 \pi^2 T_0} p^2 \,p_0^2\, f_{eq} (p) \, v\, \ln{\left( \frac{\omega - k \, v + \frac{i}{\tau_{eq}}}{\omega + k \, v + \frac{i}{\tau_{eq}}}\right)} \bigg),\no\\
    &G^{0 3, 0 0}(\omega, k) = \frac{\omega}{k} \left( G^{0 0, 0 0}(\omega, k) + \frac{\varepsilon_{eq} + P_{eq}}{c_s^2}\right) = - (\varepsilon_{eq} + P_{eq}) \, \frac{ \, i \,\tau_{eq} \, \omega \, b}{a c - b^2 c_s^2} = G^{0 0, 0 3}(\omega, k), \no\\
    &G^{0 3, 0 3}(\omega, k) = \frac{\omega}{k} G^{0 0, 0 3}(\omega, k) - \left(\varepsilon_{eq} + P_{eq}\right),\no\\
   &G^{0 3, 3 3}(\omega, k) = \frac{\omega}{k} G^{0 3, 0 3}(\omega, k),\no\\
   & G^{0 3, 1 1}(\omega, k) = G^{0 3, 2 2}(\omega, k)  = \frac{\omega}{k} \left( G^{0 0, 1 1}(\omega, k) - \left(\varepsilon_{eq} + P_{eq}\right) \right),\no\\
    & G^{0 i, 0 i}(\omega, k) = \frac{k}{\omega}G^{0 i, 3 i}(\omega, k) = \frac{- i \tau_{eq}\, k}{d} \, \int \, \frac{dp}{2 \pi^2 T_0} p^2 \,p_0^2\, f_{eq} (p) \mathcal{I}_6(v, \omega, k),\qquad i = 1, 2, \no\\
   & G^{1 2, 1 2}(\omega, k) = - i \tau_{eq} \omega \, \int \, \frac{dp}{2 \pi^2 T_0} p^2 \,p_0^2\, f_{eq} (p) \mathcal{I}_9(v, \omega, k),\no\\
   & G^{3 i, 0 i}(\omega, k) = \frac{k}{\omega}G^{3 i, 3 i}(\omega, k) = \frac{\omega}{k} G^{0 i, 0 i}(\omega, k).
\end{align}
It is evident that the Ward identity relates different two-point functions and the only independent components are $(G^{0 0, 0 0}, G^{0 0, 1 1}, G^{0 1, 0 1}, G^{1 2, 1 2})$. To proceed, we give a detailed expansion of the $b$ parameter in the small and large $x$ limits
\begin{align}\label{eq: expansion-b}
     b\bigg\vert_{x \ll 1} &= \frac{-i}{k \tau_{eq} c_s^2} \bigg\{ 
    1 + \frac{i \omega \tau_{eq} -1}{2 i k \tau_{eq}}
    \bigg(L(\omega, k) + g_1(x) \left( L(\omega, k) - \frac{2 i k \tau_{eq} (1 - i \tau_{eq} \omega)}{(1 - i \tau_{eq} \omega)^2 + k^2 \tau_{eq}^2}\right)\no\\
   & \hspace{1.75cm} - g_2(x) \left( L(\omega, k) + \frac{2 i k \tau_{eq} (1 - i \tau_{eq} \omega) \left( k^2 \tau_{eq}^2 - (1 - i \tau_{eq} \omega)^2 \right)}{\left((1 - i \tau_{eq} \omega)^2 + k^2 \tau_{eq}^2\right)^2}\right) + \mathcal{O}(\frac{x^6}{ \Tilde{p}^6})
    \bigg)
    \bigg\},\no\\
    b\bigg\vert_{x \gg 1} &= - \frac{i k \tau_{eq}}{(1 - i \tau_{eq} \omega)^2} \bigg( h_1(x) - h_2(x) \frac{k \tau_{eq}\left( 3 k^2 \tau_{eq}^2 + 5 (1 - i \tau_{eq} \omega)^2\right)}{(1 - i \tau_{eq} \omega)^2}\bigg),\no\\
\end{align}
where
\begin{align}
    & g_1(x) \equiv \frac{x \left(x K_1(x) + K_2(x)\right)}{2 \left( x K_2(x) + 3 K_3(x)\right)}, \qquad g_2(x) \equiv \frac{x^3 \left( K_2(x) - 3 K_0(x)\right)}{16 \left( x K_2(x) + 3 K_3(x)\right)},\no\\
    & h_{1}(x) \equiv \frac{5 x K_2(x) + (30 + x^2) K_3(x)}{x^2 K_3(x)}, \qquad h_2(x) \equiv \frac{7 x K_3(x) + (56 + x^2) K_4(x)}{x^3 K_3(x)}.
\end{align}
\subsection{Hydrodynamic limit}
After obtaining the correlation functions, we must match their small momentum and frequency limits to the universal hydrodynamic behaviors. The hydrodynamic limit of the no-pole correlation function $G^{1 2, 12}(\omega, k)$ for a non-conformal system up to the third-order transports is known as follows \cite{Grozdanov:2015kqa}
\begin{align}\label{eq:hydro-G1212}
    G^{1 2, 1 2}(\omega, k) &= P_{eq} - i \eta \omega + \left( \eta \tau_\pi - \frac{\kappa}{2} +\kappa^*\right) \omega^2 - \frac{\kappa}{2}   k^2 \no\\
& + \frac{i}{2}  \left( \upsilon^{(3,2)}_{31} + \upsilon^{(3,2)}_{32} - \upsilon^{(3,2)}_{34}  \right) \omega^3  + \frac{i}{2} \left(\upsilon^{(3,2)}_1 - \upsilon^{(3,2)}_{31} - \upsilon^{(3,2)}_{32}  \right) \omega k^2.
\end{align}
To obtain the transport coefficient, we should derive the expression of $G^{1 2, 1 2}(\omega, k)$ provided in Eq. \eqref{eq:corr-TT-p1}. The small $x$ expansion of this correlation function takes the following form
\begin{align}\label{eq:exp-G1212}
    &G^{1 2, 1 2}(\omega, k)\bigg\vert_{x \ll 1} =  i \tau_{eq} \omega \, \frac{\varepsilon_{eq} + P_{eq}}{8 c_s^2}\bigg\{\frac{(1-i \tau  \omega )^3 }{(k \tau )^4} +\frac{5 (1-i \tau  \omega ) c_s^2}{(k \tau )^2} + \frac{(1-i \tau  \omega ) \left(k^2 \tau ^2+(1-i \tau  \omega )^2\right)g_1(x)}{ (k \tau )^4} \no\\
    &\hspace{2cm} -\frac{(1-i \tau  \omega ) \left(7 k^2 \tau ^2+(1-i \tau  \omega )^2\right) g_2(x)}{ (k \tau )^4} + \frac{i}{2} L(\omega, k) \mathcal{G}(\omega, k, x) \bigg\},
\end{align}
where
\begin{align}
    \mathcal{G}(\omega, k, x) &\equiv \frac{\left(k^2 \tau ^2+(1-i \tau  \omega )^2\right)^2 }{ (k \tau )^5} +\frac{\left(k^2 \tau ^2+(1-i \tau  \omega )^2\right) \left(-3 k^2 \tau ^2+(1-i \tau  \omega )^2\right) g_1(x)}{ (k \tau )^5} \no\\
    & +\frac{ \left(15 k^4 \tau ^4+6 k^2 \tau ^2 (1-i \tau  \omega )^2-(1-i \tau  \omega )^4\right) g_2(x)}{4 (k \tau )^5}.
\end{align}
We perform the small $(\omega, k)$ expansion of Eq. \eqref{eq:exp-G1212} and compare it to Eq. \eqref{eq:hydro-G1212}. This shall give us transport coefficients for small $x$
\begin{align}
    &\frac{\eta}{T_0^3}  = T_0 \tau_{eq} \left(\frac{4}{5 \pi^2} - \frac{x^2}{6 \pi^2} + \mathcal{O}(x^4)\right),\no\\
    &\frac{\eta}{S_{eq}}  = T_0 \tau_{eq} \left(\frac{1}{5} - \frac{x^2}{60} + \mathcal{O}(x^4)\right),\no\\
    & \upsilon^{(3,2)}_1 - \upsilon^{(3,2)}_{31} - \upsilon^{(3,2)}_{32} = T_0^4 \tau_{eq}^3 \left(\frac{8}{35 \pi^2} - \frac{x^2}{15 \pi^2} + \mathcal{O}(x^4)\right).
\end{align}
For large $x$, the following expression is obtained
\begin{align}\label{eq:exp-G1212-large-x}
    &G^{1 2, 1 2}(\omega, k)\bigg\vert_{x \gg 1} =  i \tau_{eq} \omega \, \frac{\varepsilon_{eq} + P_{eq}}{8 c_s^2}\bigg\{\frac{(1-i \tau  \omega )^3 }{(k \tau )^4} +\frac{5 (1-i \tau  \omega ) c_s^2}{(k \tau )^2} - \frac{\mathcal{H}(\omega, k, x)}{ k^4 \tau_{eq}^4 (1 - i \tau_{eq} \omega)}\bigg\},
\end{align}
where
\begin{align}
  \mathcal{H}(\omega, k, x) \equiv (1 - i \tau_{eq} \omega)^4 + 5 k^2 \tau_{eq}^2 (1 - i \tau_{eq} \omega)^2 c_s^2 h_1(x) + k^2 \tau_{eq}^2 \left( 8 k^2 \tau_{eq}^2 - 25 (1 - i \tau_{eq} \omega)^2\right) c_s^2 h_2(x).
\end{align}
We expand the latter two-point function in small $\omega$ and $k$ and compare it with Eq. \eqref{eq:hydro-G1212} for large $x$. The result has become as follows
\begin{align}
  &\frac{\eta}{T_0^3}  = T_0 \tau_{eq} \frac{e^{-x} x^{3/2}}{2\sqrt{2} \pi^{3/2}},\no\\
    & \upsilon^{(3,2)}_1 - \upsilon^{(3,2)}_{31} - \upsilon^{(3,2)}_{32} = 0,\no\\
    &\kappa = \kappa^* = 0, \qquad \tau_\pi = \tau_{eq},\no\\
    & \upsilon^{(3,2)}_{31} + \upsilon^{(3,2)}_{32} - \upsilon^{(3,2)}_{34}  = 2 \eta \tau^2_{eq}.
\end{align}

The mass's presence explicitly breaks the conformal symmetry, as shown in section \ref{sec: thermo}. Therefore, the bulk viscosity is no longer zero, \(\xi \neq 0\). To obtain the bulk viscosity, we note the following relation \cite{Jeon:1994if}
\begin{align}\label{eq: bulk}
    \xi = - \frac{1}{9} \frac{d}{d \omega} \mbox{Im} \, G_{\tilde{P}, \tilde{P}}(\omega, k \to 0) \bigg \vert_{\omega \to 0},
\end{align}
where 
\begin{align}
    \tilde{P} \equiv  \delta T^\mu_\mu = \delta T^0_0 - \sum_{i=1}^3 \delta T^i_i.
\end{align}
According to Eqs. \eqref{eq:main-Tmunu}, we obtain
\begin{align}\label{eq: bulk-main}
    & \tilde{P} = c_0 \delta T^{0 0} + c_3 \delta T^{0 3} + \cdots,
\end{align}
where
\begin{align}
    & c_0 = 1 -  \frac{2 c_s^2}{i \tau_{eq} \omega (\varepsilon_{eq} + P_{eq})} \int \, \frac{dp}{2 \pi^2 T_0} p^2 \,p_0^2\, f_{eq} (p)  \mathcal{I}_4(v, \omega, k) - \frac{c_s^2 (1 - i \tau_{eq} \omega)}{k^2 \tau_{eq}^2 (\varepsilon_{eq} + P_{eq})} \int \, \frac{dp}{2 \pi^2 T_0} p^2 \,p_0^2\, f_{eq} (p)  \mathcal{I}_1(v, \omega, k),\no\\
    & c_3 = - \frac{2}{\varepsilon_{eq} + P_{eq}} \int \, \frac{dp}{2 \pi^2 T_0} p^2 \,p_0^2\, f_{eq} (p)  \mathcal{I}_6(v, \omega, k) - \frac{1}{\varepsilon_{eq} + P_{eq}} \int \, \frac{dp}{2 \pi^2 T_0} p^2 \,p_0^2\, f_{eq} (p)  \mathcal{I}_5(v, \omega, k).
\end{align}
Dotted terms in Eq. \eqref{eq: bulk-main} are irrelevant for computing bulk viscosity. Regarding Eq. \eqref{eq:corr-TT-p1}, the bulk viscosity is derived from the following relation
\begin{align}
   \xi = - \frac{1}{9} \frac{d}{d \omega} \mbox{Im} \, \bigg( \left(c_0 + c_3 \frac{\omega}{k}\right)^2 \,  G^{0 0, 0 0}(\omega, k = 0) \bigg)\bigg \vert_{\omega \to 0}.
\end{align}
To simplify this expression, we benefit from Eqs. \eqref{eq: somes}, \eqref{eq:corr-TT-p1}, \eqref{eq:App-B1}, and \eqref{eq:App-B2}. After some computations, we get
\begin{align}
    & \frac{\xi}{T_0^3} = \frac{T_0 \tau_{eq}}{135 c_s^2}\bigg( 4 g_1(x) \left(35 -87 c_s^2+ 15 \left(3 c_s^2-4\right) g_1(x)\right)-5 \left(3 c_s^2+4\right) - 24 g_2(x) \left(30 -77 c_s^2+ 20 \left(3 c_s^2-5\right) g_1(x)\right) \no\\
    & \hspace{2cm} + 2880 \left(c_s^2-2\right) g_2(x)^2 + \frac{c_s^2(129-48 g_1(x))}{1 - 2 g_1(x) + 8 g_2(x)}+\frac{54 c_s^2 (g_1(x)-1)}{(1 - 2 g_1(x) + 8 g_2(x))^2}\bigg). 
\end{align}
The small $x$ expansion of bulk viscosity reads as follows
\begin{align}
   & \frac{\xi}{T_0^3}\bigg\vert_{x \ll 1} = \frac{T_0 \tau_{eq}}{1620} x^4 + \mathcal{O}(x^6). 
\end{align}

\begin{figure}
    \centering
    \includegraphics[width=0.625\textwidth]{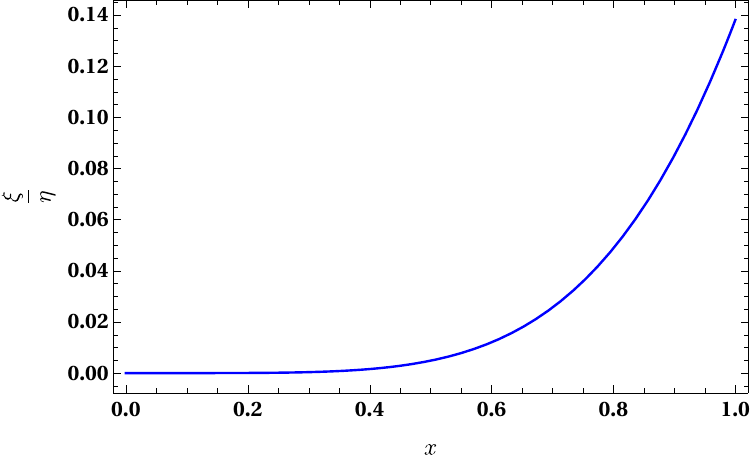}
\caption{Ratio of $\xi/\eta$ as a function of $x$.}
    \label{fig: plotxe}
\end{figure}

In what follows, we will derive the hydro modes in the sound channel. To do this, we expand the $G^{0 0, 0 0}$ for small $(\omega, k)$ that leads to the following expression for small $x$
\begin{align}\label{eq: small-G00}
    G^{0 0, 0 0}(\omega, k)\bigg \vert_{x \ll 1} = \frac{(\varepsilon_{eq} + P_{eq})\, i\, k^2 \tau_{eq} \bigg(-1 + 2 g_1(x) - 8 g_{2}(x)\bigg)}{\omega \bigg( 1 - 3 c_s^2 (1 - i \tau_{eq} \omega) - 2 g_1(x) + 8 g_2(x) \bigg)\, \mathcal{D}^{(s)}_{sound}(\omega, k, x)},
\end{align}
where
\begin{align}
    &\mathcal{D}^{(s)}_{sound}(\omega, k, x) = 1 + \frac{c_s^2 \, k^2 \, \tau_{eq} \mathcal{F}_{sound}^{(1)}(\omega, x)}{\mathcal{F}_{sound}^{(2)}(\omega, x)},\\
    & \mathcal{F}_{sound}^{(1)}(\omega, x) = -4 g_1(x) \left(40 i g_2(x)+9 \tau_{eq}  \omega +5 i\right) + 20 i g_1(x)^2 + 8 g_2(x) \left(40 i g_2(x) + 27 \tau_{eq}  \omega + 10 i\right) + 9 \tau_{eq}  \omega +5 i,\no\\
    & \mathcal{F}_{sound}^{(2)}(\omega, x) = 5 \omega  (1 - i \tau_{eq}  \omega) \left(2 g_1(x) - 8 g_2(x) - 1\right) \bigg(1 - 3 c_s^2 (1 - i \tau_{eq}  \omega ) - 2  g_1(x) + 8  g_2(x)\bigg).\no
\end{align}
Eq. \eqref{eq: small-G00} has three distinct poles at 
\begin{align}\label{eq: poles-sound-x}
& \omega_{1} = \frac{8 i k^2 \tau_{eq}}{x^6 (\gamma_E-\ln{2})} + \mathcal{O}(k^4),\no\\
& \omega_{2} = - \frac{i}{\tau_{eq}}  - i k^2 \tau_{eq} (\frac{4}{15} -\frac{x^2}{45})+ \mathcal{O}(k^4),\no\\
& \omega_{3} = \frac{i x^6 (\gamma_E-\ln{2})}{24\tau_{eq}} + \frac{8 i k^2 \tau_{eq}}{x^6 (\ln{2} - \gamma_E)} + \mathcal{O}(k^4),
\end{align}
where $\gamma_E = 0.5772$ is the Euler number. The massless limit of the pole $\omega_{2}$ continuously merges into the known result \cite{Romatschke:2015gic}, but the massless limit of the other two poles, namely $\omega_1$ and $\omega_3$, is different from the analogous result ones.  Indeed, the mass disappears the propagating sound modes and turns them into dissipative modes. Since $\gamma_E-\ln{2} \simeq - 0.115 < 0$, the lowest lying modes are stable against fluctuation.

For large $x$, the retarded two-point function has following expansion
\begin{align}
    G^{0 0, 0 0}(\omega, k)\bigg \vert_{x \gg 1} = \frac{(\varepsilon_{eq} + P_{eq})\, i\, k^2 \tau_{eq} (1 - i \tau_{eq} \omega) h_1(x)^3}{ \mathcal{D}^{(l)}_{sound}(\omega, k, x)},
\end{align}
where
\begin{align}
  \mathcal{D}^{(l)}_{sound}(\omega, k, x) = h_1(x)^3 \left( \omega - i \tau_{eq} (\omega^2 + c_s^2 k^2)\right)  -  \omega (1 - i \tau_{eq} \omega)^2 h_1(x)^2 - 5 k \tau_{eq} \omega (1 - i \tau_{eq} \omega)^2 h_2(x) \bigg(h_1(x) + 5 h_2(x) k \tau_{eq}\bigg).
\end{align}
At large $x$, the following modes exist at the lowest level
\begin{align}
    & \omega_1 = i k^2 \tau_{eq} (\frac{1}{5} + \frac{1}{x}) + \mathcal{O}(k^4),\no\\
    & \omega_2 = \frac{5 i}{\tau_{eq} x} - \frac{i k^2 \tau_{eq}}{5} + \mathcal{O}(k^4),\no\\
    & \omega_3 = - \frac{i}{\tau_{eq}} - \frac{i k^2 \tau_{eq}}{x} + \mathcal{O}(k^4).
\end{align}
Unlike the $(\omega_2, \omega_3)$ modes, which are stable, the mode $\omega_1$ is unstable in large $x$ limit.

In the shear channel, we expand $G^{0 i, 0 i}$ given in Eq. \eqref{eq:corr-TT-p1} for small momenta at small $x$ as follows
\begin{align}\label{eq:shear-mode}
    & G^{0 i, 0 i}(\omega, k) \bigg\vert_{x \ll 1} = \frac{c_s^2 \, \bigg(70 (1 - i \tau_{eq} \omega)^2\, \left(-1 + 3 c_s^2 + 2 g_1(x) - 8 g_2(x)\right) + k^2 \tau_{eq}^2 \mathcal{N}^{(s)}_{shear}(x)\bigg)}{70 (1 - i \tau_{eq} \omega)\, \bigg(1 - 3 c_s^2 (1 - i \tau_{eq} \omega)\, - 2 g_1(x) + 8 g_2(x)\bigg) \bigg( 1 - \frac{k^2 \tau_{eq}^2 \mathcal{D}^{(s)}_{shear}(\omega, x)}{6160 (1 - i \tau_{eq} \omega)^2} \bigg)},
\end{align}
where
\begin{align}
 & \mathcal{N}^{(s)}_{shear}(x) = \frac{32 - 180 c_s^2 + 8 g_1(x) \left(135 c_s^2 - 93 g_2(x) - 32\right) - 3 g_2(x) \left(780 c_s^2 + 461 g_2(x) - 112\right) + 272 g_1(x)^2}{8 g_1(x) - 33 g_2(x) - 2},\no\\
 & \mathcal{D}^{(s)}_{shear}(\omega, x) = 461 + \frac{77 \left(99 c_s^2 (1-i \tau_{eq}  \omega )+2 g_1(x)-17\right)}{1 - 3 c_s^2 (1 - i \tau_{eq}  \omega ) - 2 g_1(x) + 8 g_2(x)}+\frac{160 (7 - 94 g_1(x))}{ 8 g_1(x) - 33 g_2(x) - 2}.
\end{align}
Eq. \eqref{eq:shear-mode} has three modes at small frequencies as follows
\begin{align}
    & \omega_{1, 2} = - \frac{i}{\tau_{eq}} \pm \frac{4 k}{\sqrt{70}} \left(1 - \frac{x^2}{24} + \cdots\right)  + \mathcal{O}(k^2),\no\\
    & \omega_3 = \frac{i x^6 (\gamma_E-\ln{2})}{24\tau_{eq}} + i k^2 \tau_{eq} \left(- \frac{1}{5} + \frac{x^2}{60}\right) + \mathcal{O}(k^4).
\end{align}
The massless limit of the latter equation coincides with the analogous results in \cite{Romatschke:2015gic}. For large $x$, the corresponding retarded Green function reads as follows
\begin{align}
    & G^{0 i, 0 i}(\omega, k) \bigg\vert_{x \gg 1} = \frac{(1 - i \tau_{eq} \omega) \left(\varepsilon_{eq} + P_{eq}\right) \mathcal{N}^{(l)}_{shear}(x, \omega, k)}{\mathcal{D}^{(l)}_{shear}(x, \omega, k)},
\end{align}
where
\begin{align}
    & \mathcal{N}^{(l)}_{shear}(x, \omega, k) = (1 - i \tau_{eq} \omega)^2 (h_1(x) - 1) - \left(k^2 \tau_{eq}^2 + 5 (1 - i \tau_{eq} \omega)^2\right) h_2(x), \no\\
    & \mathcal{D}^{(l)}_{shear}(x, \omega, k) = - \mathcal{N}^{(l)}_{shear}(x, \omega, k) - i \tau_{eq} \omega (1 - i \tau_{eq} \omega)^2.
\end{align}
The lowest lying modes are solutions of $\mathcal{D}^{(l)}_{shear}(x, \omega, k) = 0$ which are
\begin{align}
 & \omega_{1, 2} = - \frac{i}{\tau_{eq}} \pm \frac{i k}{x^{1/2}} + \frac{i k^2 \tau_{eq}}{2 x} + \mathcal{O}(k^4),\no\\
    & \omega_3 = - \frac{i k^2 \tau_{eq}}{x} + \mathcal{O}(k^4).
\end{align}
All modes are stable when $x \to \infty$.
\section{Analytical Behaviors}
Investigating the relationship between the branch cut, poles, and hydrodynamic modes is a crucial objective. In Ref. \cite{Romatschke:2015gic}, it has been proposed that the hydrodynamic poles coincide with the poles of the retarded Green’s function in the strong coupling regime, i.e., $\tau_{eq} T \to 0$. However, in the weak coupling regime, i.e., $\tau_{eq} T \to \infty$, a logarithmic branch cut arises between $\omega = k$ and $\omega = -k$. In this section, we will examine this scenario for massive systems.  

To investigate this paradigm in the charge correlators, we expand the retarded two-point function of Eq. \eqref{eq:final-corr-Rcharge} in the weak coupling regime and small $x$:
\begin{align}  
    G_{0 0}^{(l)}(x; \omega, k) \bigg\vert_{\tau_{eq} T \to \infty} &= \chi(x) \bigg( 1 + \frac{\omega}{2 k} \left(L(\omega, k) + \frac{x^2 K_0(x) \left(\frac{2 k \omega  \left(k^2+\omega ^2\right)}{\left(k^2-\omega ^2\right)^2}+L(\omega, k)\right)}{8 K_2(x)}+\frac{x K_1(x) \left(L(\omega, k)-\frac{2 k \omega }{k^2-\omega ^2}\right)}{2 K_2(x)}\right)\bigg) \no\\  
   & + \mathcal{O}\left(\frac{1}{\tau_{eq}}\right).  
\end{align}  
This shows that the weak coupling limit of the two-point function exhibits a logarithmic branch cut, with its endpoints located at $\omega = k$ and $\omega = -k$, similar to the massless case. Considering this expansion, we get  
\begin{align}  
    \frac{1}{\omega} \mbox{Im}\, G_{0 0}^{(l)}(x; \omega, k=0)\bigg\vert_{\tau_{eq} T \to \infty} &= \pi \chi(x) \, \delta(\omega) \, \left( 1 + \frac{x^2 K_0(x) + 4 x K_1(x)}{8 K_2(x)}\right)\no\\
    & = \pi \chi_{x = 0} \, \delta(\omega) \, \left( 1 - \frac{x^2 }{64}\right),  
\end{align}  
which is always positive and respects the positivity condition of the spectral density. We adopt $\chi_{x = 0} = \chi(m = 0) = \frac{T_0^2}{\pi^2}$. In the strong coupling regime, the expansion of the retarded two-point function matches Eq. \eqref{eq:G00-l}, demonstrating that hydrodynamic poles emerge in this regime.  

\begin{figure}
    \centering
\includegraphics[width=0.485\textwidth,valign=t]{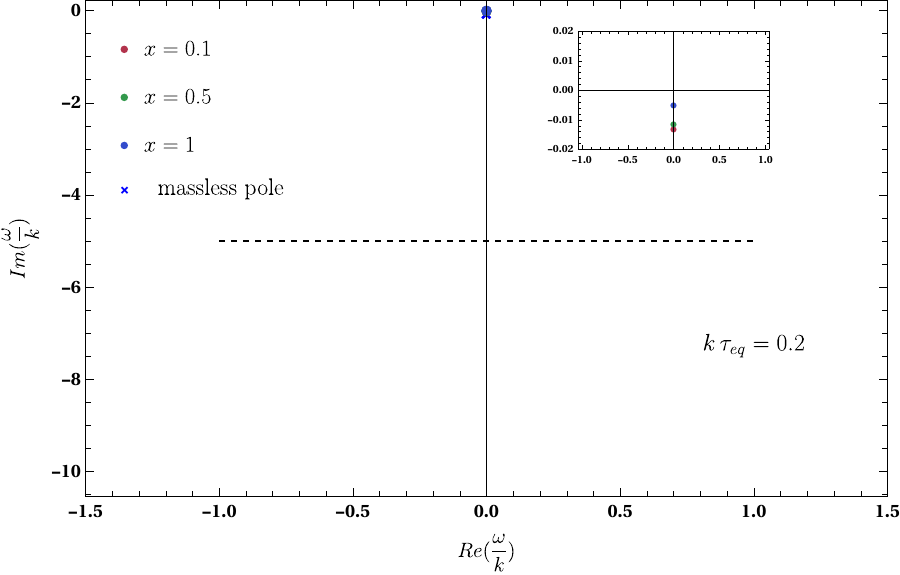}
\hspace{0.3cm}    \includegraphics[width=0.485\textwidth,valign=t]{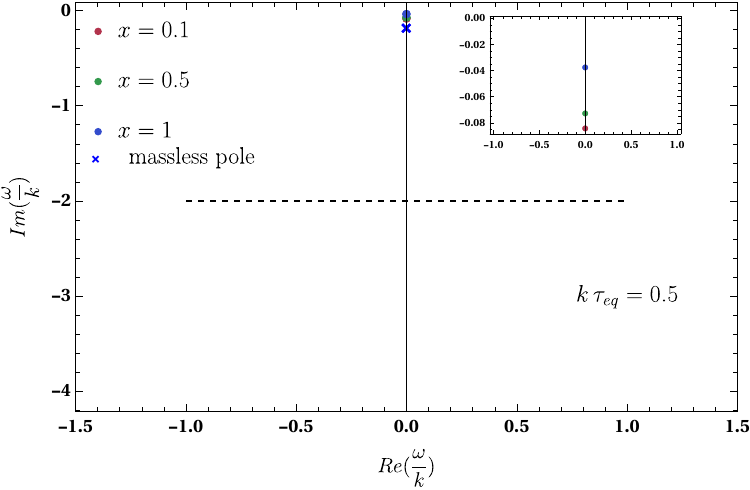}\\
    \vspace{0.3cm}
    \includegraphics[width=0.485\textwidth,valign=t]{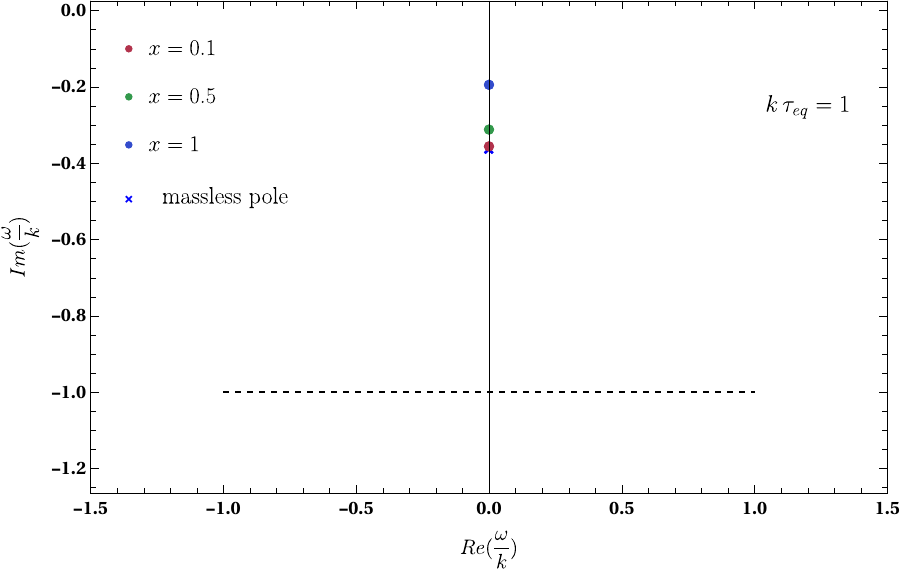}
\hspace{0.3cm}
\includegraphics[width=0.486\textwidth,valign=t]{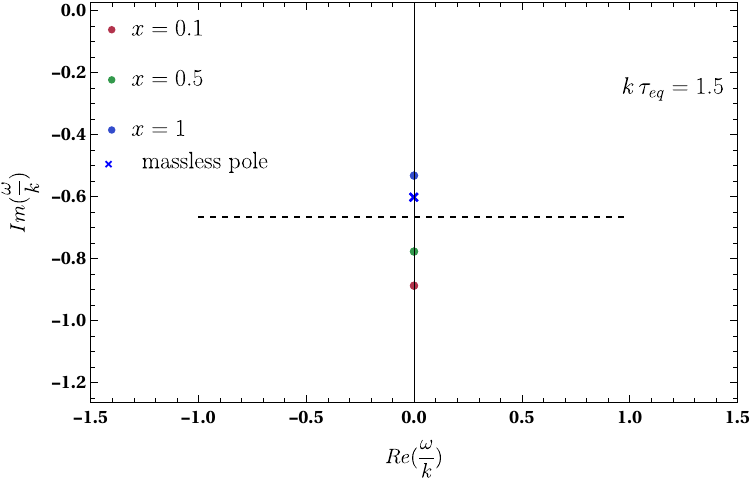}
\caption{Comparison between hydro poles for massless and massive cases in the charge correlation functions. In each figure, the dashed line indicates the location of the branch cut, and massless poles are shown with cross marks. Various "$x$" ratios are displayed in different colors. Left (right) top plots correspond to $k \tau_{eq} = 0.2 (k \tau_{eq} = 0.5)$, and left (right) bottom plots demonstrate $k \tau_{eq} = 1 (k \tau_{eq} = 1.5)$.}
    \label{fig: plot-kt-branch-cut-pole-Rcharge}
\end{figure}

For non-zero momentum, the charge correlators exhibit a logarithmic branch cut in the lower-half complex plane between $\omega = - \frac{i}{\tau_{eq}} - k$ and $\omega = - \frac{i}{\tau_{eq}} + k$. We find that for $k \tau_{eq} < \frac{\pi}{2}$, poles emerge above the branch cut, with their small $x$ expansion given by
\begin{align}  
    \omega = -\frac{i}{\tau_{eq}} + i k \cot{(k \tau_{eq})} + \frac{i k x^2}{4} \left( - \cot{(k \tau_{eq})} + \frac{k \tau_{eq}}{\sin^2{(k \tau_{eq})}}\right) + \mathcal{O}(x^4).  
\end{align}  
This solution corresponds to the small-mass limit of $\omega{(h)}$ in Eq. \eqref{eq:hydro-Rcharge-small-x} and coincides at $x=0$ with the results in \cite{Romatschke:2015gic}. In Fig. \ref{fig: plot-kt-branch-cut-pole-Rcharge}, we show the lowest-lying poles of the retarded $G_{0 0}$ correlator given in Eq. \eqref{eq:G00-l} for different $x$ values in the strong coupling regime at various $k \tau_{eq}$. Dashed lines indicate the positions of the branch cuts, and the hydrodynamic poles of the massless system are marked with crosses. Similar to Fig. \ref{fig: plot-x-kt}, the mass tends to shift the poles toward the unstable region.  

\begin{figure}
    \centering
    \includegraphics[width=0.485\textwidth]{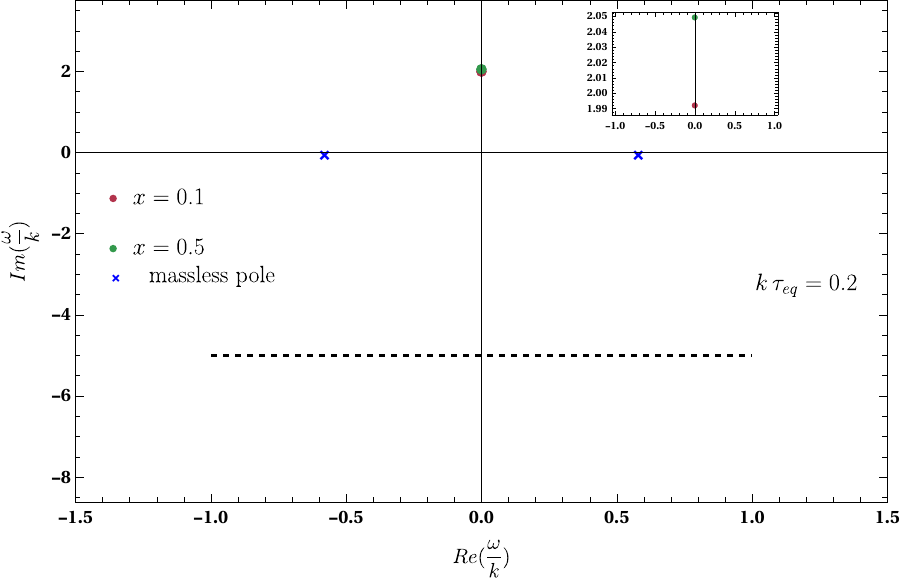}
\hspace{0.3cm}
    \includegraphics[width=0.485\textwidth]{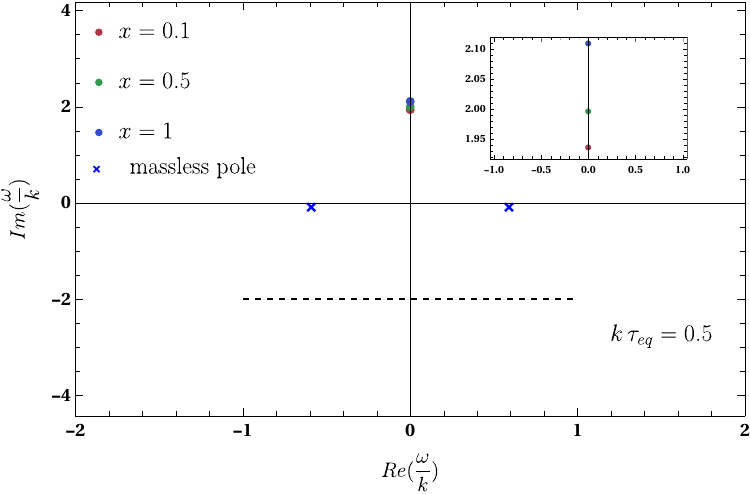}\\
    \vspace{0.3cm}
    \includegraphics[width=0.485\textwidth]{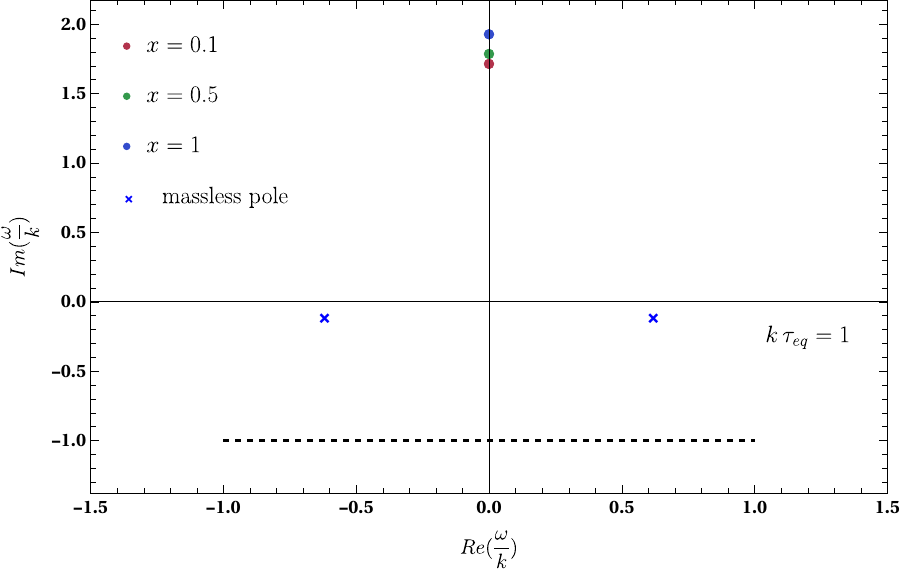}
\hspace{0.3cm}
    \includegraphics[width=0.485\textwidth]{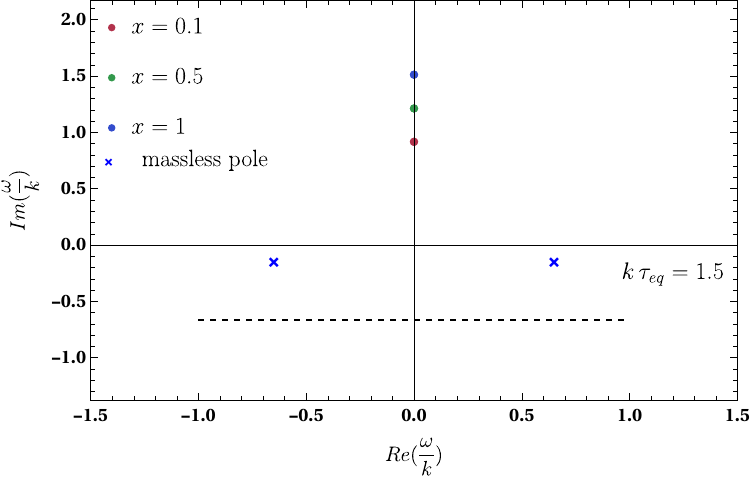}
\caption{Comparison between hydro poles for massless and massive cases in the sound channel. Conventions are similar to Fig. \ref{fig: plot-kt-branch-cut-pole-Rcharge}.}
    \label{fig: plot-kt-branch-cut-pole-EM}
\end{figure}

\begin{figure}
    \centering
    \includegraphics[width=0.485\textwidth]{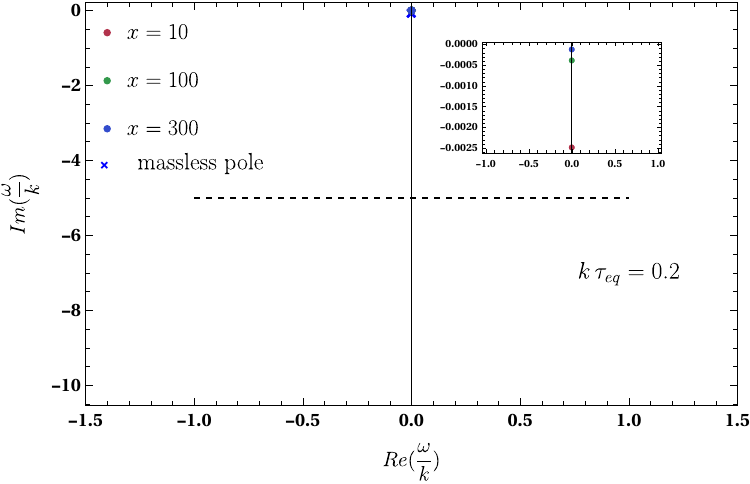}
\hspace{0.3cm}
    \includegraphics[width=0.485\textwidth]{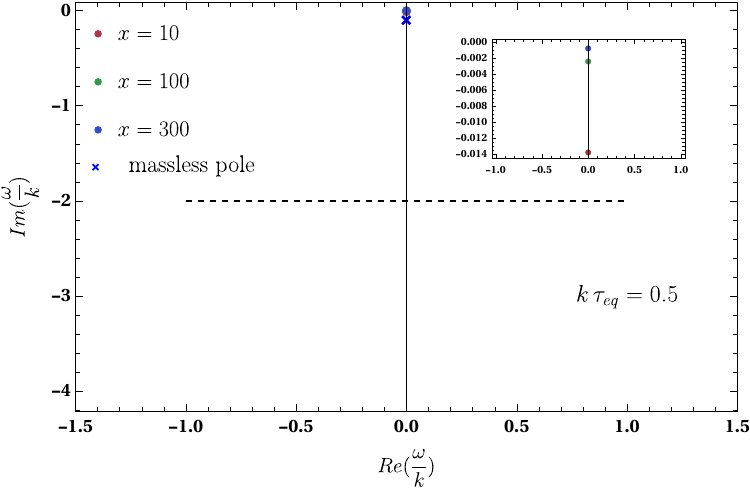}\\
    \vspace{0.3cm}
    \includegraphics[width=0.485\textwidth]{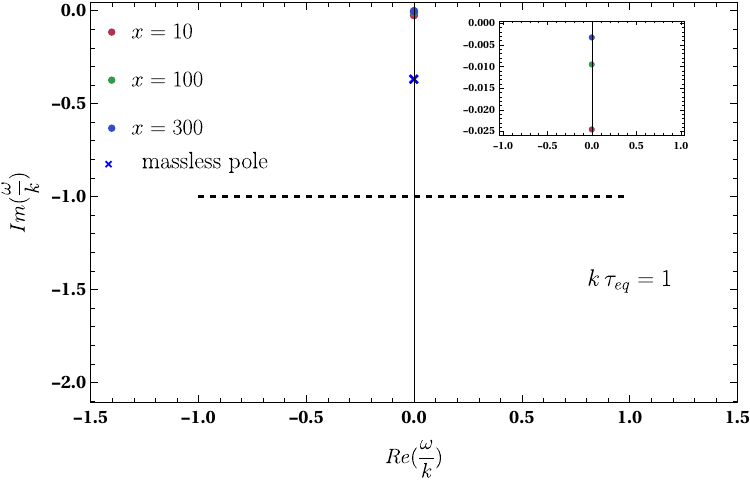}
\hspace{0.3cm}
    \includegraphics[width=0.485\textwidth]{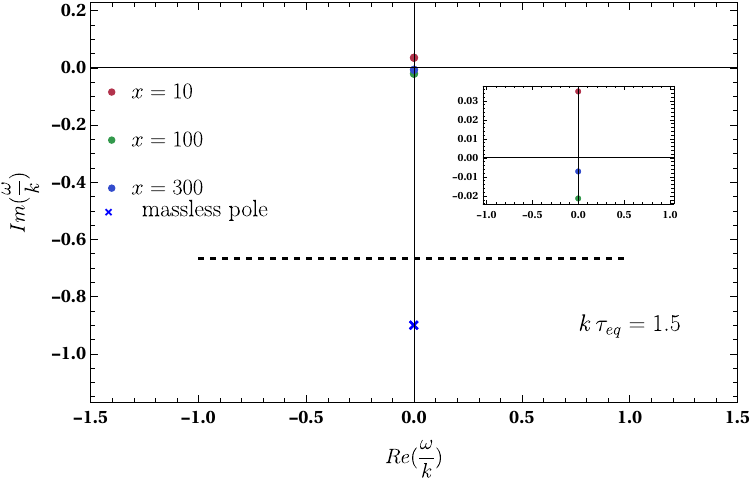}
\caption{Sketch of poles for large massive particles in the charge correlation function. Massless poles are indicated with a cross sign. Inside plots show the precise location of poles for $x = (10, 100, 300)$.}
    \label{fig: plot-kt-branch-cut-pole-CC-LX}
\end{figure}

\begin{figure}
    \centering
    \includegraphics[width=0.485\textwidth]{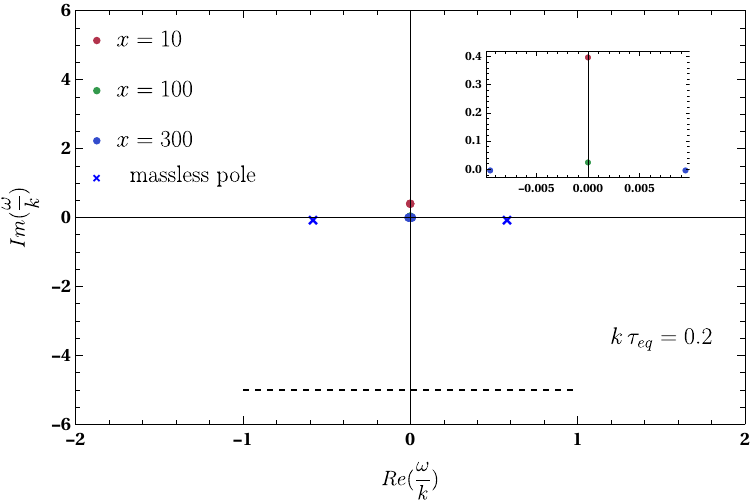}
\hspace{0.3cm}
    \includegraphics[width=0.485\textwidth]{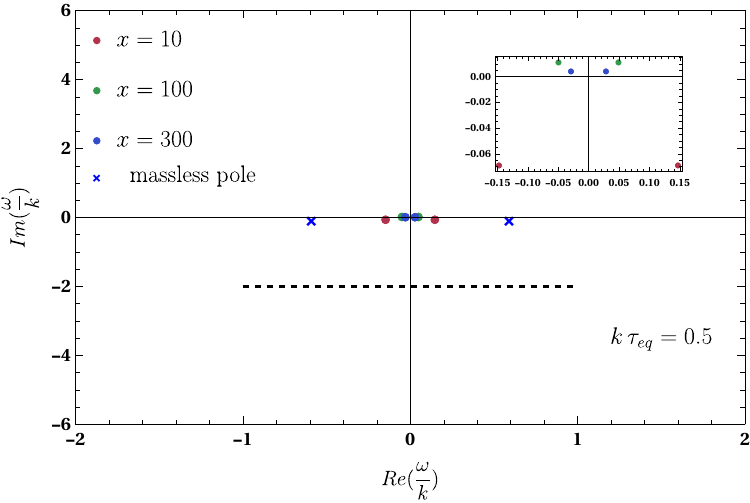}\\
    \vspace{0.3cm}
    \includegraphics[width=0.485\textwidth]{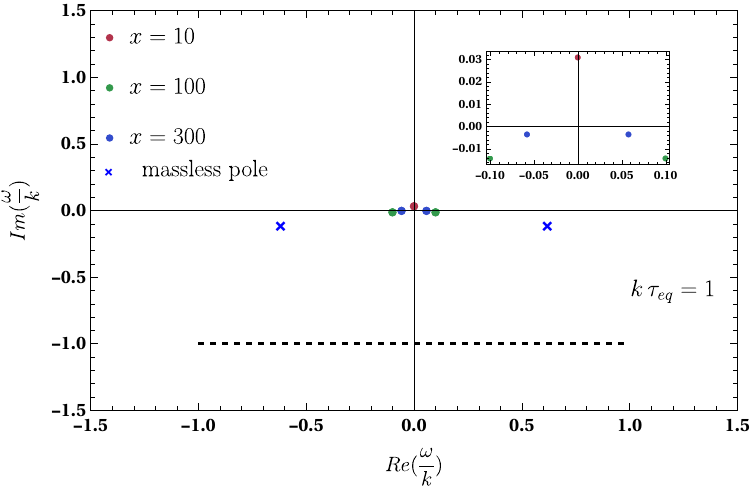}
\hspace{0.3cm}
    \includegraphics[width=0.485\textwidth]{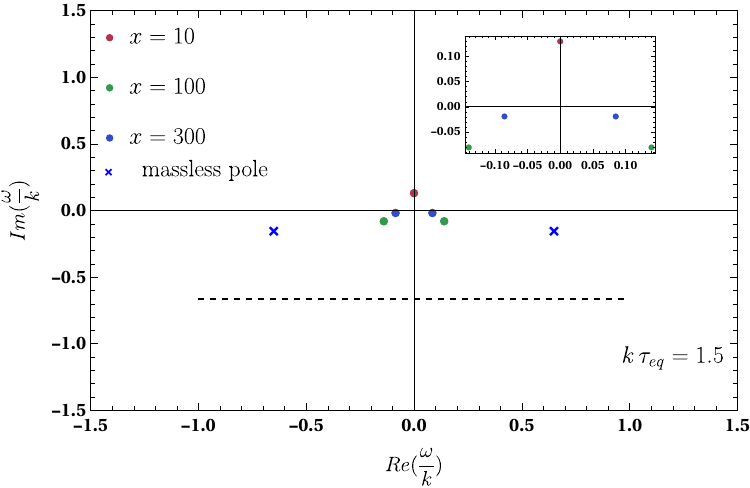}
\caption{Structure of poles for large massive particles $x = (10, 100, 300)$ in the sound channel. Insets realize the accurate positions of poles for different $x$.}
    \label{fig: plot-kt-branch-cut-pole-SM-LX}
\end{figure}

For the energy-momentum tensor correlators in the weak coupling limit, we obtain  
\begin{align}  
    G^{0 0, 0 0}(\omega , k) \bigg \vert_{\tau_{eq} T \to \infty} & = - \frac{\varepsilon_{eq} + P_{eq}}{c_s^2} \bigg\{ 2 + \frac{\omega}{2 k} \left( L(\omega, k) + g_1(x) \left(L(\omega, k) - \frac{2 k \omega }{k^2-\omega ^2}\right) - g_2(x) \left(\frac{2 k \omega  \left(k^2+\omega ^2\right)}{\left(k^2-\omega ^2\right)^2} + L(\omega, k)\right)\right) \no\\  
    &+ \mathcal{O}\left(\frac{1}{\tau_{eq}}\right)\bigg\}.  
\end{align}  
Similar to the R-charge correlators, a logarithmic branch cut exists between $\omega = k$ and $\omega = -k$. Additionally,  
\begin{align}  
    \frac{-1}{\omega} \mbox{Im}\, G^{0 0, 0 0}( \omega, k=0)\bigg\vert_{\tau_{eq} T \to \infty} = \frac{\varepsilon_{eq} + P_{eq}}{c_s^2} \, \pi \, \delta(\omega) \, \left( 1 + g_1(x) - g_2(x)\right) = \frac{\varepsilon_{eq} + P_{eq}}{c_s^2} \, \pi \, \delta(\omega) \, \left( 1 + \frac{x^2}{24}\right), 
\end{align}  
which satisfies the positive spectral density condition. In the strong coupling limit, the correlator expansion coincides with Eq. \eqref{eq: small-G00}, and hydrodynamic poles naturally emerge. In Fig. \ref{fig: plot-kt-branch-cut-pole-EM}, we depict the lowest-lying poles of $G^{0 0, 0 0}$ for different $x$ values in the strong coupling regime at various $k \tau$. Unlike the massless case, all poles are located on the vertical line in the unstable region. These numerical solutions agree with those given in Eq. \eqref{eq: poles-sound-x}. To analyze the results in the large $x$ limit, we present Figs. \ref{fig: plot-kt-branch-cut-pole-CC-LX} and \ref{fig: plot-kt-branch-cut-pole-SM-LX}. In Fig. \ref{fig: plot-kt-branch-cut-pole-CC-LX}, we examine the pole structure of $G_{0 0}(x; \omega, k)$ for different values of $k \tau_{eq}$, each corresponding to poles at $x = (10, 100, 300)$. The insets precisely depict the pole locations for various $x$. As momentum increases, the poles shift toward positive imaginary values, indicating an instability at high momenta. Similarly, Fig. \ref{fig: plot-kt-branch-cut-pole-SM-LX} illustrates the analytic structure of $G^{0 0, 0 0}(\omega, k)$. Here, the poles appear in pairs with opposite real parts, characteristic of propagating modes.
\section{Conclusion}
Kinetic theory provides a robust framework for macroscopic and microscopic properties of many-body systems. This capability stems from the Boltzmann equation, which treats particles as probabilistic entities, allowing for the incorporation of every relevant detail of the system. The central probabilistic variable is the one-particle distribution function, defined as the number of particles within a given space-time and momentum interval. The Boltzmann equation governs the dynamics of this function.  Upon solving the Boltzmann equation, the macroscopic properties of the system can be derived from specific moments of the distribution function. Notably, kinetic theory serves as an alternative approach to solving the intricate Navier-Stokes equations. 

With this advantage in mind, we examine the collective behavior of a system with fixed mass values. To the best of our knowledge, this work represents the first attempt to extend prior studies to the case of a massive relativistic plasma. Within the framework of Relaxation Time Approximation (RTA) kinetic theory and assuming an equilibrium state described by the Maxwell-Boltzmann distribution, we solve the Boltzmann equation for small fluctuating perturbations. This approach permits us to derive the two-point retarded correlation functions using the variational formalism. These retarded two-point functions are crucial for studying bulk properties, as they contain essential physical insights. For instance, functions with poles yield valuable information about propagating modes, while transport coefficients can be explicitly determined by comparing the low-frequency limit of these functions with their counterparts in general hydrodynamic relations. 
In the current-current correlation functions, we obtain the collective modes as well as the diffusion constant for both small and large $x$ values.
In the energy-momentum sector of perturbations, we have computed the complete spectrum of retarded Green’s functions and verified the fulfillment of the Ward identity. The relevant transport coefficients have been derived up to third order in $(\omega, k)$ for extreme $x$ limits. A key observation in the sound channel is that all hydrodynamic modes are purely imaginary. In contrast, the structure of hydrodynamic modes in the shear channel (comprising non-hydrodynamic and diffusive modes) remains unchanged, though the mass modifies the coefficients in the dispersion relation. Importantly, the mass preserves the branch-cut structure, and for \(k \tau_{eq} \leq \pi/2\), the hydrodynamic modes lie above the branch cut in the strong coupling regime, indicating the emergence of collective modes. 

This study underscores the critical role of mass in relativistic plasmas. Several promising extensions of this work can be pursued. For instance, investigating temperature or energy-dependent mass profiles with an appropriate equation of state could provide deeper insights into the role of mass in more physical scenarios. This is valuable because near the phase transition points, mass serves as an order parameter running with temperature or any other parameters. Given the significance of mass effects near critical points, this framework could be instrumental in studying the QCD phase diagram, where hydrodynamic instabilities may play a crucial role. We follow the strategy of ``$x$'' expansion, but one can perform the integrals numerically and examine the collective behavior at each mass-to-temperature value. 
\begin{appendix}
    \section{Useful Integrals}\label{sec-integrals} 
    In this part, we are going to provide helpful integral relations that were used in previous sections. The following relations can be derived easily \cite{jeffrey2008handbook}
    \begin{align}
    &K_n(x) = \frac{\sqrt{\pi}}{\Gamma(n+\frac{1}{2})} \, (\frac{x}{2})^n \, \int_1^\infty dz \, (z^2 -1)^{n-\frac{1}{2}} \, e^{- z x},\no\\
    & K_{n+1}(x) = \frac{\sqrt{\pi}}{\Gamma(n+\frac{1}{2})} \, (\frac{x}{2})^n \, \int_1^\infty dz \, z\, (z^2 -1)^{n-\frac{1}{2}} \, e^{- z x},\no\\
    &\left(- \frac{1}{x}\frac{\partial}{\partial x}\right)^m \left(x^n K_{n}(x)\right) =  x^{n-m} K_{n-m}(x), \qquad \left(- \frac{1}{x}\frac{\partial}{\partial x}\right)^m \left(\frac{K_{n}(x)}{x^n}\right) = \frac{K_{n+m}(x)}{x^{n+m}},\no\\
    & K_{n+1}(x) = K_{n-1}(x) + \frac{2 n}{x} K_n(x),\qquad \frac{\partial K_n(x)}{\partial x} = - \frac{K_{n-1}(x) + K_{n+1}(x)}{2}.
\end{align}
Afterward, we obtain the following results
\begin{align}
    & \int \, \frac{dp}{2 \pi^2 T_0} p^6 \, f_{eq} (p) = \frac{15 T_0^6}{2 \pi^2} x^4 K_4(x), \no\\
    & \int \, \frac{dp}{2 \pi^2 T_0} p^4 \, f_{eq} (p) = \frac{3 T_0^4}{2 \pi^2} x^3 K_3(x), \no\\
    &\int \, \frac{dp}{2 \pi^2 T_0} p^2 \, f_{eq} (p) = \frac{T_0^2}{2 \pi^2} x^2 K_2(x),\no\\
    & \int \, \frac{dp}{2 \pi^2 T_0} p^6 \,p_0^2 \, f_{eq} (p) = \frac{ 15 T_0^8}{2 \pi^2} x^4 \left(7 x K_3(x) + (56 + x^2) K_4(x)\right),\no\\
     & \int \, \frac{dp}{2 \pi^2 T_0} p^4 \,p_0^2 \, f_{eq} (p) = \frac{ 3 T_0^6}{2 \pi^2} x^3 \left(5 x K_2(x) + (30 + x^2) K_3(x)\right),\no\\
    & \int \, \frac{dp}{2 \pi^2 T_0} p^2 \,p_0^2 \, f_{eq} (p) = \frac{ T_0^4}{2 \pi^2} x^3 \left(x K_2(x) + 3 K_3(x)\right),\no\\
    &  \int \, \frac{dp}{2 \pi^2 T_0} p^2 \, p_0 \, f_{eq} (p) = \frac{T_0^3}{2 \pi^2} x^2 \left(x K_1(x) + 3 K_2(x)\right),\no\\
    & \int \, \frac{dp}{2 \pi^2 T_0}  \, f_{eq} (p) = \frac{1}{2 \pi^2} x K_1(x),\no\\
    & \int \, \frac{dp}{2 \pi^2 T_0} \, p_0^2\, f_{eq} (p) = \frac{ T_0^2}{2 \pi^2} x^2 \left(x K_1(x) +  K_2(x)\right),
    \end{align}
where $x = m/T_0$ and $p_0 = \sqrt{p^2 + m^2}$. Likewise, we benefit from the following integrals
    \begin{align}
    & \int \, \frac{dp}{2 \pi^2 T_0} \, \frac{p^4}{p_0}\, f_{eq} (p) = \frac{3 T_0^3}{2 \pi^2}  x^2 K_2(x), \no\\
    &\int \, \frac{dp}{2 \pi^2 T_0} \, \frac{p^2}{p_0}\, f_{eq} (p) = \frac{T_0}{2 \pi^2}  x K_1(x),\no\\ 
    & \int \, \frac{dp}{2 \pi^2 T_0} \, \frac{1}{p^2}\, f_{eq} (p) = -\frac{1}{2 \pi^2 T_0^2}  K_0(x), \no\\
    &\int \, \frac{dp}{2 \pi^2 T_0} \, \frac{1}{p_0}\, f_{eq} (p) = \frac{1}{2 \pi^2 T_0}  K_0(x), \no\\
    & \int \, \frac{dp}{2 \pi^2 T_0} \, \frac{1}{p^4}\, f_{eq} (p) = \frac{1}{6 \pi^2 T_0^4 x}  K_1(x),\no\\
    & \int \, \frac{dp}{2 \pi^2 T_0} \, \frac{p_0^2}{p^2}\, f_{eq} (p) = \frac{ x^2}{4 \pi^2} \left( K_2(x) - 3  K_0(x)\right),
\end{align}
\section{Definition of variables given in the energy-momentum sector}\label{sec-definitions}
In this part, we provide the unknown variables and functions given in Eqs. \eqref{eq: corr-charge} and \eqref{eq:main-Tmunu}. Parameters $(a, b, c, d)$ are defined as follows
\begin{align}\label{eq:App-B1}
    & a \equiv 1 - \frac{c_s^2}{2 i k \tau_{eq} (\varepsilon_{eq} + P_{eq})} \int \, \frac{dp}{2 \pi^2 T_0} p^2 \,p_0^2\, f_{eq} (p) \frac{1}{ v} \ln{\left( \frac{\omega - k \, v + \frac{i}{\tau_{eq}}}{\omega + k \, v + \frac{i}{\tau_{eq}}}\right)},\no\\
    & b \equiv \frac{1}{ i k \tau_{eq} (\varepsilon_{eq} + P_{eq})} \int \, \frac{dp}{2 \pi^2 T_0} p^2 \,p_0^2\, f_{eq} (p) \mathcal{I}_1(v, \omega, k), \no\\
    & c \equiv 1 - \frac{1 - i \tau_{eq} \omega}{ k^2 \tau_{eq}^2 (\varepsilon_{eq} + P_{eq})} \int \, \frac{dp}{2 \pi^2 T_0} p^2 \,p_0^2\, f_{eq} (p) \mathcal{I}_1(v, \omega, k), \no\\
    & d \equiv 1 - \frac{1}{ i \tau_{eq} \omega (\varepsilon_{eq} + P_{eq})} \int \, \frac{dp}{2 \pi^2 T_0} p^2 \,p_0^2\, f_{eq} (p) \mathcal{I}_4(v, \omega, k).
    \end{align}
    Also, the other variables are listed below
    \begin{align}\label{eq:App-B2}
    & \mathcal{E}_0 \equiv  - \tau_{eq} \int \, \frac{dp}{2 \pi^2 T_0} p^2 \,p_0^2\, f_{eq} (p) \bigg\{ \Gamma^0_{0 0}(\omega, \Vec{k}) \frac{1}{2 i k \tau_{eq} v} \ln{\left( \frac{\omega - k \, v + \frac{i}{\tau_{eq}}}{\omega + k \, v + \frac{i}{\tau_{eq}}}\right)}  \no\\
    & \hspace{1cm} + 2 \Gamma^0_{0 3}(\omega, \Vec{k}) \frac{1}{ i k \tau_{eq} } \mathcal{I}_1(v, \omega, k) + \Gamma^0_{3 3}(\omega, \Vec{k}) \frac{1 - i \tau_{eq} \omega}{ k^2 \tau_{eq}^2} \mathcal{I}_1(v, \omega, k) + (\Gamma^0_{1 1}(\omega, \Vec{k}) + \Gamma^0_{2 2}(\omega, \Vec{k})) \frac{\mathcal{I}_4(v, \omega, k)}{i \tau_{eq} \omega}\bigg\}, \no\\
    & \mathcal{E}_3 \equiv - \tau_{eq} \int \, \frac{dp}{2 \pi^2 T_0} p^2 \,p_0^2\, f_{eq} (p) \bigg\{ \Gamma^0_{0 0}(\omega, \Vec{k}) \frac{1}{ i k \tau_{eq} } \mathcal{I}_1(v, \omega, k)  + \Gamma^0_{3 3}(\omega, \Vec{k}) \mathcal{I}_5(v, \omega, k) \no\\
    & \hspace{1cm} + 2 \Gamma^0_{0 3}(\omega, \Vec{k}) \frac{1 - i \tau_{eq} \omega}{  k^2 \tau_{eq}^2 } \mathcal{I}_1(v, \omega, k)  + (\Gamma^0_{1 1}(\omega, \Vec{k}) + \Gamma^0_{2 2}(\omega, \Vec{k})) 
    \mathcal{I}_6(v, \omega, k)\bigg\}, 
    \end{align}
   where $\mathcal{I}_i(v, \omega, k)$ functions for $(i = 1, \cdots, 9)$ are defined in below
    \begin{align}
       &\mathcal{I}_1(v, \omega, k) \equiv  \int \frac{d\Omega}{4\pi} \frac{i\,  \tau_{eq} \Vec{v}(\Vec{p}) \cdot \Vec{k}}{1 + \tau_{eq} (- i \omega + i \Vec{k} \cdot \Vec{v}(\Vec{p}))} = 1 + \frac{i\, \omega \, \tau_{eq} -1}{2 i \, \tau_{eq}\, k \, v}\,\ln{\left( \frac{\omega - k \, v + \frac{i}{\tau_{eq}}}{\omega + k \, v + \frac{i}{\tau_{eq}}}\right)},\no\\
 & \mathcal{I}_2(v, \omega, k) \equiv   \int \frac{d\Omega}{4\pi} \frac{v_z(\Vec{p})}{1 + \tau_{eq} (- i \omega + i \Vec{k} \cdot \Vec{v}(\Vec{p}))} = - \frac{i}{k\, \tau_{eq}} \mathcal{I}_1(v, \omega, k),\no\\
   &\mathcal{I}_3(v, \omega, k) \equiv   \int \frac{d\Omega}{4\pi} \frac{i\, \tau_{eq}\, \omega\, v_z^2(\Vec{p})}{1 + \tau_{eq} (- i \omega + i \Vec{k} \cdot \Vec{v}(\Vec{p}))} = \frac{i\, \omega (1-i \tau_{eq} \, \omega)}{k^2 \, \tau_{eq}}\, \mathcal{I}_1(v, \omega, k),\no\\
   &\mathcal{I}_4(v, \omega, k) \equiv \int \frac{d\Omega}{4\pi} \frac{i\, \tau_{eq}\, \omega\, v_x^2(\Vec{p})}{1 + \tau_{eq} (- i \omega + i \Vec{k} \cdot \Vec{v}(\Vec{p}))} = - \frac{i \tau_{eq} \omega}{2} \left( \frac{1 - i \tau_{eq} \omega}{(\tau_{eq}\, k )^2} + \frac{(1 - i \tau_{eq} \omega)^2 + \tau_{eq}^2 k^2 v^2}{2 v  (i \tau_{eq}\, k )^3} \ln{\left( \frac{\omega - k \, v + \frac{i}{\tau_{eq}}}{\omega + k \, v + \frac{i}{\tau_{eq}}}\right)}\right).\no\\
    &\mathcal{I}_5(v, \omega, k) \equiv  \int \frac{d\Omega}{4\pi} \frac{ v_z^3(\Vec{p})}{1 + \tau_{eq} (- i \omega + i \Vec{k} \cdot \Vec{v}(\Vec{p}))} = \frac{ (1 - i \tau_{eq} \omega)^2}{(i k \tau_{eq})^3} + \frac{ v^2}{3 i k \tau_{eq}} - \frac{(1 - i \tau_{eq} \omega)^3}{ 2 v k^4 \tau_{eq}^4} \ln{\left( \frac{\omega - k \, v + \frac{i}{\tau_{eq}}}{\omega + k \, v + \frac{i}{\tau_{eq}}}\right)}, \no\\
    &\mathcal{I}_6(v, \omega, k) \equiv  \int \frac{d\Omega}{4\pi} \frac{ v_z(\Vec{p}) v_x^2(\Vec{p})}{1 + \tau_{eq} (- i \omega + i \Vec{k} \cdot \Vec{v}(\Vec{p}))}  = \int \frac{d\Omega}{4\pi} \frac{ v_z(\Vec{p}) v_y^2(\Vec{p})}{1 + \tau_{eq} (- i \omega + i \Vec{k} \cdot \Vec{v}(\Vec{p}))} \no\\
    & \hspace{1.5cm} = -\frac{(1 - i \tau_{eq} \omega)^2}{ 2 (i k \tau_{eq})^3} + \frac{ v^2}{3 i k \tau_{eq}} + \frac{(1 - i \tau_{eq} \omega) \left( (1 - i \tau_{eq} \omega)^2 + v^2 k^2 \tau_{eq}^2\right)}{4 v k^4 \tau_{eq}^4} \ln{\left( \frac{\omega - k \, v + \frac{i}{\tau_{eq}}}{\omega + k \, v + \frac{i}{\tau_{eq}}}\right)},\no\\
    &\mathcal{I}_7(v, \omega, k) \equiv  \int \frac{d\Omega}{4\pi} \frac{ v_z^4(\Vec{p}) }{1 + \tau_{eq} (- i \omega + i \Vec{k} \cdot \Vec{v}(\Vec{p}))}  \no\\
    & \hspace{1.5cm} = -\frac{ (1 - i \tau_{eq} \omega)^3}{ (k \tau_{eq})^4} + \frac{ v^2 (1 - i \tau_{eq} \omega)}{3  k^2 \tau_{eq}^2} + \frac{(1 - i \tau_{eq} \omega)^4 }{ 2 v (i k \tau_{eq})^5} \ln{\left( \frac{\omega - k \, v + \frac{i}{\tau_{eq}}}{\omega + k \, v + \frac{i}{\tau_{eq}}}\right)},\no\\
    &\mathcal{I}_8(v, \omega, k) \equiv  \int \frac{d\Omega}{4\pi} \frac{ v_z^2(\Vec{p}) v_x^2(\Vec{p})}{1 + \tau_{eq} (- i \omega + i \Vec{k} \cdot \Vec{v}(\Vec{p}))}  = \int \frac{d\Omega}{4\pi} \frac{ v_z^2(\Vec{p}) v_y^2(\Vec{p})}{1 + \tau_{eq} (- i \omega + i \Vec{k} \cdot \Vec{v}(\Vec{p}))} \no\\
    & \hspace{1.5cm} = \frac{(1 - i \tau_{eq} \omega)^3}{2 (k \tau_{eq})^4} + \frac{v^2 (1 - i \tau_{eq} \omega)}{3  k^2 \tau_{eq}^2} - \frac{(1 - i \tau_{eq} \omega)^2 \left( (1 - i \tau_{eq} \omega)^2 + v^2 k^2 \tau_{eq}^2\right)}{4 v (i k \tau_{eq})^5} \ln{\left( \frac{\omega - k \, v + \frac{i}{\tau_{eq}}}{\omega + k \, v + \frac{i}{\tau_{eq}}}\right)},\no\\
    &\mathcal{I}_9(v, \omega, k) \equiv  \int \frac{d\Omega}{4\pi} \frac{  v_x^2(\Vec{p}) v_y^2(\Vec{p})}{1 + \tau_{eq} (- i \omega + i \Vec{k} \cdot \Vec{v}(\Vec{p}))}  \no\\
    & \hspace{1.5cm} = - \frac{(1 - i \tau_{eq} \omega)^3}{8 (k \tau_{eq})^4} - \frac{5 v^2 (1 - i \tau_{eq} \omega)}{24  k^2 \tau_{eq}^2} + \frac{ \left( (1 - i \tau_{eq} \omega)^2 + v^2 k^2 \tau_{eq}^2\right)^2}{16 v (i k \tau_{eq})^5} \ln{\left( \frac{\omega - k \, v + \frac{i}{\tau_{eq}}}{\omega + k \, v + \frac{i}{\tau_{eq}}}\right)},
\end{align}
Here, we define $v_z = v \cos{\theta}, v_x = v \sin{\theta} \cos{\phi}, $ and $v_y = v \sin{\theta} \sin{\phi}$. 
\end{appendix}

\bibliographystyle{fullsort}
\bibliography{Refs}
\end{document}